\documentclass[twoside]{article}
\include{ijmpd}
\include{epsf}

\newcommand{\beq}{\begin{equation}}
\newcommand{\eeq}{\end{equation}}

\def \({\left( }
\def \){\right) }
\newcommand{\beqar}{\begin{eqnarray}}
\newcommand{\eeqar}{\end{eqnarray}}

\newcommand{\I}{\mbox{\rm i}}

\begin{document}
\setlength{\textheight}{7.7truein}    

\runninghead{Evolution equations...} {Stavridis \& Kokkotas}
\normalsize\textlineskip
\thispagestyle{empty} \setcounter{page}{1}

\copyrightheading{}     

\vspace*{0.88truein}

\fpage{1} \centerline{\bf EVOLUTION EQUATIONS FOR SLOWLY ROTATING
STARS} \vspace*{0.37truein}

\centerline{\footnotesize ADAMANTIOS STAVRIDIS}
\vspace*{0.015truein}
\centerline{ \footnotesize \it Departimento di Physica, University
of Rome, "La Sapienza", Rome, Italy}
\centerline{ \footnotesize \it Department of Physics, Aristotle
University of Thessaloniki, 54124 Greece}
\centerline
{\footnotesize KOSTAS D. KOKKOTAS}
\vspace*{0.015truein}
\centerline{ \footnotesize \it Department of Physics, Aristotle
University of Thessaloniki, 54124 Greece}
\publisher{(received date)}{(revised date)}

\vspace*{0.21truein}
\abstract{
  We present a hyperbolic formulation of the evolution equations
  describing non-radial perturbations of slowly rotating relativistic stars
  in the Regge--Wheeler gauge.
  We demonstrate the stability preperties of the new evolution set of equations and compute the polar w-modes for slowly rotating stars.
  }

\section{Introduction}
The theory of non-radial perturbations of relativistic stars has
been a field of intensive study for almost four decades, beginning
with the pioneering paper of Thorne \& Campolattaro\cite{TC67} in
the mid-60s followed by a sequence of papers by Thorne and
collaborators. The authors focused mainly on perturbations of
non-rotating relativistic stars. The study lasted for over three
decades and today we are quite confident that we have a good
understanding of the oscillation spectrum of non-rotating
relativistic stars\cite{KS99}. The study of rotating relativistic
stars is a considerably more elaborate project. The construction
of a rotating stellar background is by itself a very difficult
task and during the last two decades a number of numerical codes
were written for this purpose\cite{Stergioulas2003}. The study of
the response of a rotating star to small perturbations is also a
very difficult analytic and numerical task. Fast rotating stars
are not spherically symmetric objects and the decomposition of the
perturbations in spherical harmonics, which can in principle
simplify the problem, is usually extremely difficult if not
impossible. A way to study perturbations of rotating stars is to
expand the perturbation equations in powers of the rotational
parameter $\epsilon=\Omega/\Omega_K$, where $\Omega$ is the
angular velocity of the star and $\Omega_K$ is the angular
velocity at the mass shedding limit. In this way, and up to 2nd
order in $\epsilon$, the decomposition into spherical harmonics of
the perturbation equations is feasible and the perturbation
functions will be characterized by only one temporal and one
spatial coordinate i.e. ($t,r$). This ``slow rotation
approximation'' is a very good tool in  studying typical compact
stars. For example, for the fastest known rotating pulsar with
period of 1.56ms, we get $\epsilon \sim 0.3$. Yet compact stars
during their birth could be rotating even faster and  become
dynamically or secularly unstable. For such cases and for the
exact estimation of the critical angular velocity for the onset of
these instabilities one might consider to do better than the
``slow rotation'' approximation and abolish the use of spherical
harmonics. Then, due to the axisymmety of the background star, the
perturbation functions will be characterized by two spatial
coordinates $(r,\theta)$ and the temporal one $t$. Finally, with
the advance of computational power it is now possible to evolve
the full set of non-linear Einstein's equations describing the
rotating body and from there to extract the characteristic
frequencies of oscillation\cite{FSK2000,FDGS2001,SAF2004}. Still
in both cases i.e. the ``fast rotation" and the``non-linear
evolution" the extraction of the various features of the modes is
quite cumbersome if not impossible without the proper
understanding of these features in the ``slow rotation"
approximation. This article serves this purpose, it presents a
system of perturbation equations in the slow rotation
approximation, which are stable and can be used to study the
fundamental features of the rotating star perturbations.

The full set of perturbation equations for rotating relativistic
stars is quite complicated. During the last decade we got the
first results for oscillation modes of rotating relativistic
stars\cite{Kojima93b,LF99,lo99,YE2001,LAF2001,RSK03,LFA2003}, but
there is still a long way to go before we have a full
understanding of the behaviour of the oscillation spectrum. In
most studies mentioned earlier the slow-rotation approximation was
used to tackle the problem. There are typically two methods to
study the spectrum of the oscillation frequencies. The first
method is by Fourier transforming the equations, where one drops
the time dependence of the equations and the system can be solved
as a boundary value problem, given that the correct boundary
conditions have been provided. The second approach is the
evolution of the system of linear (or non-linear) perturbation
equations and then by Fourier transforming the perturbation
functions one can get the characteristic frequencies which happen
to be excited for a given set of initial data. Both techniques
have their own advantages and disadvantages, for example in the
first approach the correct definition of the boundary conditions
is vital for the correct estimation of the frequencies, but this
might be difficult for fast rotating compact stars. On the other
hand via this method one can get the complete spectrum of
oscillation modes. The time evolution technique does not depend
critically on accurate definition of the boundary conditions but
the excitation of specific modes depends on the initial data
provided. For a generic initial perturbation one should get all
modes of the coupled spectrum, but they may be difficult to be
found if excited at small amplitude or due to the finite evolution
time. This means that one can reveal via this method only specific
parts of the oscillation spectrum.

Chandrasekhar \& Ferrari\cite{CF91} were the first to study
axisymmetric perturbations in the slow rotation approximation and
apart from deriving the system of equations they have also showed
how rotation induces coupling of the polar and axial modes, which
actually decouple for non-rotating stars \footnote{Polar or
spheroidal or even parity modes are characterized by a sign change
under parity transformation according to $(-1)^l$, while the axial
or toroidal or odd parity ones change as $(-1)^{l+1}$.}. Soon
afterwards, Kojima\cite{k92} presented the first complete
derivation of the coupled polar and axial perturbation equations.
But, in the specific form that they have been given, the equations
where not extremely useful for time-evolutions. Kojima used the
well-known Regge-Wheeler gauge for the spacetime perturbations.
Allen et~al.\cite{Allen98} managed to derive from Kojima's set of
equations a set of two coupled wave equations which together with
the appropriate constraints could describe the perturbations of a
{\em non-rotating} background. Finally, Ruoff et~al.\cite{RSK02}
derived an alternative set of evolution equations in a gauge
suggested thirty years ago by Battiston et~al\cite{BCL71}. In
their derivation Ruoff et~al.\cite{RSK02} used the ADM form of
linearized Einstein's equations\cite{Ruoff2001} and showed that
the use of the Regge-Wheeler gauge leads to unnecessary terms with
second order spatial derivatives of the lapse function $\alpha$
and the 3-spatial metric $\gamma_{ij}$. The Battiston et.al. (BCL)
gauge,  appears to be more suitable for the linearized ADM
formulation of Einstein's equations but it seems to suffer from
numerical instabilities which have not yet been resolved. It is
not actually clear whether the numerical instabilities observed
are due to the known instability of the ADM formulation of
Einstein's equations or it is due to a combination of the
formulation and the choice of gauge\cite{ADM_inst}.

This is the reason that we derived a system of hyperbolic
equations describing the linear perturbations of slowly rotating
stars using the standard form of Einstein's equations (not the ADM
formulation) in the commonly used Regge-Wheeler gauge. These
equations reduce, in the non-rotating case, directly to the
equations of Allen et~al.\cite{Allen98} (polar perturbations) and
in the slow-rotating case to the axial ones that have been used by
Andersson\cite{Nils98} and Ruoff \& Kokkotas \cite{RK01,RK02} in
the study of the relativistic r-mode spectrum.

As a test of stability of the present system, we have derived
a reduced version of the equations describing only the spacetime perturbations in
the so called inverse Cowling Approximation (ICA)\cite{AKS96}.
The fluid part is actually unaffected by the choice of gauge.
We show that in the present form the system of evolution equations is stable.
Additionally, we have calculated, for the first time, the polar
w-modes for rotating stars.

In the next parts of this section we briefly describe the
equations for the construction of the background stellar model and
the derivation of the perturbation equations. In the second
section, we show the form of the evolution equations for both
fluid and spacetime perturbations. The derivation of the evolution
equations for the fluid perturbations, in a format compatible with
the equations describing the spacetime perturbations is given in
Appendix I. For clarity, we have chosen to give the lengthy
background coefficients of the evolution equations in Appendix II
at the end. In the third section, we describe a simplified form of
the evolution equations for the spacetime, i.e. we ``freeze" the
fluid perturbations and keep only the part of the system that
deals with the spacetime equations (ICA). We evolve this system of
equations as a test of stability and compute the
w-modes\cite{KS92} for some typical rotating stellar models.
Finally, in the last section we discuss the analytic and numerical
results that have been presented.

\subsection{The background model}

Here we will briefly present the basic equations describing the
stationary configuration. We consider a uniformly rotating
relativistic star with angular velocity $\Omega$, which is
assumed to be {\em small} compared to $\Omega_K$, the angular
velocity of the mass shedding limit. We will keep terms of first
order in $\Omega$ only. The parameter that we will use to expand
all the rotational quantities will be $\epsilon=\Omega/\Omega_K$.
Higher order terms in the construction of the background models as
well as in the perturbation equations will be omitted.

Under the above assumptions the background metric of a slowly
rotating spherically symmetric star has the form
 \begin{eqnarray}
ds^2&=&g_{\mu\nu}^{(0)}dx^{\mu} dx^{\nu}
\nonumber \\
&=&-e^{\nu}dt^2+e^{\lambda}dr^2+r^2(d\theta^2+\sin^2\theta
d\phi^2)-2\omega r^2 \sin^2\theta d t d\phi,
 \end{eqnarray}
where $\nu$ and $\lambda$ are functions of the radial coordinate
$r$. The function $\omega=\omega(r)$ describes the dragging of the
inertial frames and is a first-order rotational correction.

We will also assume a perfect fluid described by the energy-momentum
tensor:
\begin{equation}
T_{\mu \nu}= (\rho +p)u_\mu u_\nu + p g_{\mu \nu}
\end{equation}
with a 4-velocity vector of the form
\begin{equation}
u_{\mu}= \left(e^{-\nu/2},\ 0, \ 0, \ \Omega e^{-\nu/2}\right)
\end{equation}
where $p$ and $\rho$ are the pressure and the energy density. The
static spherical stellar configuration will be constructed using
the Tolman-Oppenheimer-Volkof (TOV) equations\cite{BFS}. The
solution of the TOV equations provides the metric functions
(potentials) $\nu=\nu(r)$ and $\lambda=\lambda(r)$ and the
remaining function $\omega=\omega(r)$ (dragging) will be
calculated in the stellar interior via the equation\cite{Hartle67}:
\begin{equation}
\omega''-\left[\kappa(\rho+p)r e^\lambda +4 r^{-1}\right]\omega' -
4k(\rho+p)e^\lambda \omega=0
\end{equation}
where $k=4\pi$. Outside the star, $\omega$ gets the simple form
$\omega=2J/r^3$, where $J$ is the angular momentum of the star. An additional
function
\begin{equation}
\varpi \equiv \Omega-\omega \ .
\end{equation}
will be used in various equations for simplicity reasons. Note,
that in the present study we will use a barotropic form of the
equation of state i.e. $p=p(\rho)$ , which serves well for the
problems that we consider here.

\subsection{The perturbed configuration}

For the derivation of the perturbation equations we assume a a
metric perturbation $h_{\mu \nu}$ which is small compared to the
background metric $g_{\mu\nu}^{(0)}$ and the perturbed spacetime
will be described by a metric of the form
\begin{equation}
g_{\mu\nu}=g_{\mu\nu}^{(0)}+h_{\mu\nu}\ .
\end{equation}
Still we will use for lowering and raising the indices the
background metric $g_{\mu\nu}^{0}$.

For the fluid perturbations we assume a small velocity
perturbation $\delta u^\mu$ and small density ($\delta \rho$) and
pressure ($\delta p$) variations. With these assumptions the
perturbed Einstein equations
\begin{equation}
\delta G_{\mu \nu} =2k \delta T_{\mu \nu}
\label{eq:pert_Ein}
\end{equation}
together with the equation of motion for the fluid perturbations
\begin{equation}
\delta \left({T^{\mu \nu}}_{;\mu}\right)=0
\label{eq:pert_fluid}
\end{equation}
provide a complete set of equations for the study of small fluid
and spacetime perturbations in a spherically symmetric, slowly and
uniformly rotating background.

The components of the perturbation tensor $h_{\mu\nu}$ will be
functions of all four coordinates ($t,r,\theta,\phi$) but due to
the spherical symmetry one can expand its components in tensor
spherical harmonics.
Then by making an appropriate choice of
gauge, here we choose the Regge-Wheeler one, $h_{\mu\nu}$ gets the
form:
\begin{equation}
 h_{\mu\nu} =
 \sum_{l=2}^{\infty} \sum_{m=-l}^{l}\left(
 \begin{array}{cccc}
 e^{\nu} H_{0,lm}  &  H_{1,lm} & - h_{0,lm} {\sin^{-1}\theta}
 \partial_{\phi}
 & h_{0,lm} \sin\theta \, \partial_{\theta} \\
 \ast & e^{\lambda} H_{2,lm} &- h_{1,lm} {\sin^{-1}\theta} \partial_{\phi}
 & h_{1,lm} \sin\theta \, \partial_{\theta} \\
 \ast & \ast & r^2 K_{lm} & 0 \\
 \ast & \ast & \ast & r^2 \sin^2\theta K_{lm} \\
 \end{array}
 \right) Y_{lm}
 \end{equation}
where $H_0(t,r)$, $H_1(t,r)$, $H_2(t,r)$ and $K(t,r)$ are the
functions describing the polar perturbations, $h_0(t,r)$,
$h_1(t,r)$ describe the axial ones and the star symbol (*) stands
for the spherically symmetric components of $h_{\mu\nu}$.

The contravariant form of the perturbed fluid
velocity is\cite{k92}:
\begin{eqnarray} \delta u^t &=&\sum_{l=2}^{\infty} \sum_{m=-l}^{l}\left(\frac{1}{2}e^{-\nu/2}H_{0,lm} +\Omega
e^{-3\nu/2}h_{0,lm}\sin \theta \partial_\theta\right) Y_{l m} \nonumber\\
&+& \frac{\varpi
e^{-\nu/2}}{k(p+\rho)}\sum_{l=2}^{\infty} \sum_{m=-l}^{l}\left(V_{lm}\partial_\phi+U_{lm}\sin\theta
\partial_\theta \right)Y_{lm} \\
\delta u^r &=&  \frac{e^{\nu/2-\lambda}}{k(p+\rho)} \sum_{l=2}^{\infty} \sum_{m=-l}^{l}R_{lm}Y_{lm}\\
\delta u^\theta &=& \frac{e^{\nu/2}}{k(p+\rho)r^2}
\sum_{l=2}^{\infty} \sum_{m=-l}^{l}\left(V_{lm} \partial_\theta  - U_{lm} \sin^{-1}\theta \partial_\phi \right)Y_{lm}\\
\delta u^\phi &=&\frac{e^{\nu/2}}{k(p+\rho)r^2\sin^2\theta}
\sum_{l=2}^{\infty} \sum_{m=-l}^{l}\left(V_{lm}\partial_\phi  + U_{lm}\sin\theta\partial_\theta \right)Y_{lm}
\end{eqnarray}
where again $R(t,r)$, $V(t,r)$ are polar perturbation functions
and $U(t,r)$ describes the axial fluid perturbation. Finally, for
the density and pressure perturbations we will use the conventions
\begin{equation}
\rho_{1,lm}=\sum_{lm}\delta \rho(t,r) Y_{lm} \qquad
p_{1,lm}=\sum_{lm}\delta p(t,r) Y_{lm}
\end{equation}
For the barotropic case that we study, the sound speed relates the
variations of pressure and density i.e.
\begin{equation}
C_s^2:\equiv \frac{dp}{d\rho}\equiv \frac{\delta p}{\delta \rho} \ .
\end{equation}
Concluding, our task will be to derive a system of evolution
equations which will involve the five spacetime perturbation
functions $(H_0,H_1,H_2,K)_{lm}$, ($h_0,h_1)_{lm}$, the three velocity
perturbations $(R,V,U)_{lm}$ and the density perturbation $\rho_{1,lm}$.
There are various ways of writing the system, for example one can
derive a system of evolution equations for every single one of the
above functions.
Another way will be to evolve a certain number of
the functions and derive the rest from the constraint equations.
Here we choose the first approach since at least in the
non-rotating case it has been proven to be numerically more
stable\cite{Allen98}.

\section{Perturbation equations}

The generic form of the perturbed Einstein equations in the
background of a slowly rotating star i.e. the analytic form of
equation (\ref{eq:pert_Ein}) is actually given by
Kojima\cite{k92}. We derive in Appendix I the equations for the
fluid perturbations i.e. from equation (\ref{eq:pert_fluid}) which
were not derived by Kojima. Then we perform the appropriate
transformations in order to derive a system of first order
evolution equations for the spacetime and fluid perturbation
functions.

The difficulties in the derivation of the evolution equations and
in their numerical evolution comes from the rotation effects which
dramatically affect the structure of the system and  as a
consequence the oscillation spectra. Rotation affects the spectrum
in various ways. First, it erases the azimuthal degeneracy with
respect to $m$, second it adds rotational corrections to the
equations describing perturbations of non-rotating configuration
and third it  couples the perturbation equations of a specific
harmonic index $l$ with the $l\pm 1$ perturbations, i.e. it
couples the polar to the axial perturbations and vice-versa. In
this way the quadrupole ($l=2$) polar (spheroidal or even parity)
perturbations couple with the octupole ($l=3$) and dipole ($l=1$)
axial (toroidal or odd parity) perturbations\footnote{This is a
true if we keep rotational terms up to first order in $\epsilon$,
the inclusion of higher order terms makes the coupling more
involved i.e. it adds coupling terms of $l\pm 2$}. Therefore the
system of evolution equations, keeping rotational terms up to 1st
order in $\epsilon$ and for a specific harmonic index $l$ will be
of the following form:
\begin{eqnarray}
{\dot {\bar P}}_{lm} &=& \left({\hat \alpha_P}^{(0)}+{\hat \alpha_P}^{(1)}\right) {\bar P'}_{lm}+ \left({\hat \beta_P}^{(0)}+{\hat \beta_P}^{(1)}\right){\bar P}_{lm}
\nonumber \\
&+&{\hat \gamma_A}^{(1)}\left({{\cal L}_1^\pm},{{\cal L}_2^\pm},{{\cal L}_3^\pm}\right){\bar A'}_{lm}
+{\hat \delta_A}^{(1)}\left({{\cal L}_1^\pm},{{\cal L}_2^\pm},{{\cal L}_3^\pm}\right){\bar A}_{lm}   \label{eq:polar1} \\
{\dot {\bar A}}_{lm} &=& \left({\hat \alpha_A}^{(0)}+{\hat \alpha_A}^{(1)}\right) {\bar A'}_{lm}+ \left({\hat \beta_A}^{(0)}+{\hat \beta_A}^{(1)}\right){\bar A}_{lm}
\nonumber  \\
&+& {\hat \gamma_P}^{(1)}\left({{\cal L}_1^\pm},{{\cal L}_2^\pm},{{\cal L}_3^\pm}\right){\bar P'}_{lm}
   +{\hat \delta_P}^{(1)}\left({{\cal L}_1^\pm},{{\cal L}_2^\pm},{{\cal L}_3^\pm}\right){\bar P}_{lm} \label{eq:axial1}
\end{eqnarray}
where the dot corresponds to the time derivative and the prime to
the spatial derivative with respect to the radial coordinate. The
arrays ${\bar P}_{lm}=(K_{lm}, H_{0,lm}, H_{1,lm}, H_{2,lm},
R_{lm}, V_{lm})$ and ${\bar A}_{lm}=(h_{1,lm}, h_{0,lm}, U_{lm})$
stand for the functions which describe polar and
axial  perturbations correspondingly. The coefficient
matrices of the form ${\hat \alpha_P}^{(0)}$, ${\hat
\alpha_A}^{(0)}$,  ${\hat \beta_P}^{(0)}$ and ${\hat
\beta_A}^{(0)}$ include only background terms without rotational
corrections, while the matrices of the form ${\hat
\alpha_P}^{(1)}$, ${\hat \alpha_A}^{(1)}$, ${\hat \beta_P}^{(1)}$
and ${\hat \beta_P}^{(1)}$ are the ones which include the
rotational corrections. Finally, the matrices ${\hat
\gamma_P}^{(1)}$, ${\hat \gamma_A}^{(1)}$, ${\hat \delta_P}^{(1)}$
and ${\hat \delta_P}^{(1)}$ include background rotational terms
and especially the operators ${{\cal L}_1^\pm}$, ${{\cal
L}_2^\pm}$ and ${{\cal L}_3^\pm}$ which are  responsible for the
coupling with the $l \pm 1$ perturbations. The analytic form of
these three operators is\cite{RSK02}:
\begin{eqnarray}
{{\cal L}_1^\pm} A_{lm}&=& (l-1)Q_{lm}A_{l-1 \ m} - (l+2)Q_{l+1 \ m}A_{l+1 \ m} \\
{{\cal L}_2^\pm} A_{lm}&=&-(l+1)Q_{lm}A_{l-1 \ m} + l   Q_{l+1 \ m}A_{l+1 \ m} \\
{{\cal L}_3^\pm} A_{lm}&=&(l+1)(l-1)Q_{lm}A_{l-1 \ m} + l(l+2)Q_{l+1 \ m}A_{l+1 \ m}
\end{eqnarray}
with
\begin{equation}
Q_{lm} := \sqrt{\frac{(l-m)(l+m)}{(2l-1)(2l+1)}} \label{eq:Qlm}
\end{equation}

We should stress that because of the coupling between the
$l$-polar perturbations with the $l\pm 1$-axial and vice versa the
above sets of equations (\ref{eq:polar1})-(\ref{eq:axial1}) have
to be considered as infinite system with $l$ running from $|m|$ to
infinity, when $m$ is considered to be fixed. There is no coupling
to equations with $l<|m|$, since from its definition
(\ref{eq:Qlm}) it follows that the relevant coupling coefficient
$Q_{mm}$ is zero. Furthermore, the
equations form two independent sets, each belonging to a different
parity.  In the non-rotating case one usually distinguishes
between polar and axial perturbations.
In the rotating case, however, the polar and
axial equations are coupled, and therefore the distinction between
polar and axial modes can no longer be upheld.  However, the
equations do not mix the overall parity, which can be seen as
follows. A polar equation with even $l=|m|$ has even parity. It is
coupled to an axial equation with $l+1$, whose parity is also
even.  This axial equation in turn couples to a polar equation of
order $l+2$, which, again, has even parity. As this continues in
the same manner, we can see that for even $m$, the complete
coupled system with a leading polar equation has even parity.
Conversely, the other system starting with a leading axial
equation has odd parity. For odd $m$ we obtain the reversed
situation.  This implies that for any given $m$, it makes sense to
distinguish the modes according to their overall parity.  Lockitch
and Friedman\cite{LF99} introduced the notion of polar or axial
led modes, depending on whether the leading equations with $l=|m|$
are axial or polar.  For even $m$ the polar led modes have even
parity and the axial modes odd parity, and vice versa for odd $m$.
The axisymmetric case $m=0$ is somewhat special in that only the
even parity modes start with $l=0$ whereas the odd parity modes
have to start with $l=1$, for there are no axial $l=0$ equations.

In the derivation of the wave equations describing the spheroidal
(polar) oscillations of non-rotating stars Allen et
al.\cite{Allen98} had derived a set of two wave equations for the following combination of the metric perturbations
\begin{equation}
F_{lm}=rK_{lm} , \quad S_{lm}= \frac{e^\nu}{r} \left( H_{0,lm}-K_{lm}\right)
\end{equation}
while the enthalpy
\begin{equation}
H_{lm}= \frac{\delta p_{lm}}{\rho+p}
\end{equation}
was calculated for every time step by solving an elliptic
equation, the Hamiltonian constraint.  They have observed that the
evolution was more stable when they were using evolution equations
for all the perturbation functions $F$, $S$ and $H$. This is the
reason that in our derivation of the equations we choose to
produce an system of evolution equations for  all the function
describing both the fluid and the spacetime perturbations. A
similar set of evolution equations for non-rotating stars derived
recently by Nagar et.al.\cite{Nagar2004}, while for gauge
invariant formulations one can refer to a series of articles by
Gundlach and Martin-Garchia
\cite{Carsten2000,Carsten2001,Carsten2002}

Using a similar approach to the one used by Allen et al.\cite{Allen98}
we derived a set of wave equations for the rotational case.
The form of the evolution equations for the spacetime
perturbations is:
\begin{eqnarray}
{\ddot F} - e^{\nu-\lambda}F'' &=& a_0(F,S,H,F')+\I m
a_1\left(R,V,H_1,V',H_1',{\dot F},{\dot S}\right) \nonumber \\
 &+& {{\cal
    L}^\pm_i}a_2\left(h_0,h_1,U,h_0',h_1',U',{\dot h}_1\right)
\\
{\ddot S} - e^{\nu-\lambda}S'' &=& b_0(S,F,S')+\I m
b_1\left(R,V,H_1,R',V',{\dot H},{\dot F},{\dot S}, \ddot{V}\right) \nonumber \\
 &+& {{\cal
    L}^\pm_i}b_2\left(h_0,h_1,U,h_0',h_1',U',{\dot h}_0,{\dot h}_1,
{\dot h}_0',{\dot h}_1'\right)
\\
{\dot H}_1  &=& c_0(S,F,S')+\I m
c_1\left(R,V,H_1,{\dot F},{\dot F}'\right) \nonumber \\
 &+& {{\cal L}^\pm_i}c_2\left(h_0,U,h_0',{\dot h}_0,{\dot h}_1\right)
\\
{\ddot h}_1 - e^{\nu-\lambda}h_1'' &=& d_0(h_1,h_1')
+ \I m d_1\left(h_0,h_1,U,h_0',h_1',U',{\dot h}_1\right) \nonumber \\
&+&{{\cal L}^\pm_i} d_2\left(R,V,H_1,V',{\dot F}\right)
\\
{\dot h}_0 &=& e_0(h_1,h_1')
+ \I m e_1\left(h_0,U,h_0',{\dot h}_1\right) \nonumber \\
&+&{{\cal L}^\pm_i} d_2\left(V,{\dot F}\right)
\end{eqnarray}
Where we have used the following rule in naming the various
functional forms which are involved  in the above set of
equations. The ones of the form $a_{0}$, $b_{0}$, $c_{0}$, $d_{0}$
and $e_{0}$ do not include rotational corrections i.e. the
coefficients of the perturbations functions are of zeroth order
background terms. Those of the form $a_{1}$, $b_{1}$, $c_{1}$,
$d_1$ and $e_1$ include rotational corrections and related to the
perturbation functions which have the same parity as the evolution
function. Finally, those of the form $a_{2}$, $b_{2}$, $c_{2}$,
$d_{2}$, $e_{2}$ introduce rotational corrections and couple with
the perturbation functions of the other parity i.e. with spherical
harmonic indices $l \pm 1$. This last set of coefficients includes
the operators ${{\cal L}_1^\pm}$, ${{\cal L}_2^\pm}$ and ${{\cal
L}_3^\pm}$ which are  the ones that introduce the coupling with
the $l \pm 1$ perturbation functions.

\subsection{The first order system of equations}

It is evident that the above system of equation is not in a form
suitable for time evolutions since a significant number of
equations involve time derivatives or even mixed derivatives on
the right hand side. These terms are due to the rotational
corrections of the equations and they were not present in the
non-rotational case. We will abandon this writing (2nd order in
space and time wherever possible) and instead we will derive a
first order in time hyperbolic system of equations. Still the
above system of equations is useful in its present form for time
evolutions of the perturbations of non-rotating relativistic
stars.

We proceed by defining the new functions
\begin{eqnarray}
&&F_1={\dot F} , \quad S_1={\dot S} , \quad T_1={\dot h_1} \nonumber \\
&&F_2= F'      , \quad S_2= S'      , \quad T_2 = h_1'
\label{eq:evol_1}
\end{eqnarray}
This additional set of functions combined with appropriate
substitutions provides the new set of evolution equations. The
equations describing the evolution of the spacetime perturbations
will be:
\begin{eqnarray}
{\dot F_1} &=& e^{\nu-\lambda}F_2'
             + a_{01}F
             + a_{02} S+a_{03}H
             + a_{04} F_2
\nonumber \\
&+&a_{10}R +a_{11} V+a_{12} H_1+a_{13}V'+a_{14} H_1' + a_{15}F_1+ a_{16}S_1
\nonumber \\
&+& a_{20}h_0 + a_{21}h_1 +a_{22}U+a_{23}h_0'+a_{24}h_1'+a_{25}U'
+a_{26} {T_1} \label{eq:evol_F1}\\
\nonumber \\
{\dot S_1}
&=& e^{\nu-\lambda}S_2'
  + b_{01}S
  + b_{02} F
  + b_{03} S_2    \nonumber \\
&+& b_{10}R
  + b_{11}V
  + b_{12}H_1
  + b_{13}R'
  + b_{14}V'
  + b_{15}H_1'
  + b_{16}F_1
  + b_{17}S_1 \nonumber \\
&+& b_{20}h_0
  + b_{21}h_1
  + b_{22}U
  + b_{23}h_0'
  + b_{24}h_1'
  + b_{25}U'
  + b_{26}T_1  \nonumber \\
&+& b_{27} T_1'
  + b_{28} T_2'  \label{eq:evol_S1}\\
 \nonumber \\
{\dot H_1}&=& c_{00} F   +
              c_{01} S   +
              c_{02} S_2 \nonumber \\
          &+& c_{10} H_1 +
              c_{11} V   +
              c_{12} R +
              c_{13} F_1 +
              c_{14} F_1' \nonumber \\
          &+& c_{20} h_0 +
              c_{21} U   +
              c_{22} h_0' +
              c_{23} T_1 +
              c_{24} h_1' +  
	      c_{25} h_1
\label{eq:evol_H1}\\
\nonumber \\
{\dot T_1}&=& e^{\nu-\lambda}T_2' + d_{01} h_1 + d_{02} T_2 \nonumber \\
          &+& d_{10} h_0 + d_{11} h_1 + d_{12} U + d_{13} h_0' +
              d_{14} T_2 + d_{15} U'  + d_{16} T_1 \nonumber \\
          &+& d_{20} H_1 + d_{21} V   + d_{22} R + d_{23} V' +
              d_{24} F_1 \label{eq:evol_T1}\\
\nonumber \\
 {\dot h_0}&=& e_{00} h_1
             + e_{01} h_1'
             + e_{10} h_0
             + e_{11} U
             + e_{12} h_0'
             + e_{13} T_1 \nonumber \\
           &+& e_{20} V
             + e_{21}F_1
\label{eq:evol_h0}
\end{eqnarray}

The derivation of the fluid perturbation equations is presented in
Appendix A. The equations derived there can be combined
appropriately with the above system equations to provide a system
of four evolution equations for the perturbation of the enthalpy
function $H_{lm}$ and the three velocity perturbation functions
$R_{lm}$, $V_{lm}$ and $U_{lm}$.
They actually get the following form:
\begin{eqnarray}
{\dot H}&=&
            k_{01} R
          + k_{02} V
          + k_{03} F_1
          + k_{04} S_1
          + k_{05} H_1'
          + k_{06} R'
          + k_{07} H_1 \nonumber \\
        &+& k_{10} H
          + k_{11} S
          + k_{12} F 
	  + k_{13} F_2
	  + k_{14} F_2' \nonumber \\
        &+& k_{21} h_1
          + k_{22} T_2
\label{eq:evol_H}
\\
\nonumber \\
{\dot R}&=& f_{00} F'
          + f_{01} S'
          + f_{02} H'
          + f_{03} F
          + f_{04} S
                      \nonumber \\
        &+& f_{10} H_1
          + f_{11} V
          + f_{12} R
          + f_{13} F_1
          + f_{14} F_1'
          + f_{15} V'  \nonumber\\
        &+& {f}_{20} h_0'
          + {f}_{21} h_0
          + {f}_{22} U
          + {f}_{23} U'
          + {f}_{24} h_1
          + {f}_{25} h_1'
          + {f}_{26} T_1 \label{eq:evol_R}   \\
\nonumber \\
{\dot V}&=& q_{00} H
          + q_{01} F 
	  + q_{02} S \nonumber  \\
        &+& q_{10} H_1
          + q_{11} H_1'
          + q_{12} V
          + q_{13} R
          + q_{14} F_1
          + q_{15} S_1
          + q_{16} R' \nonumber\\
        &+& q_{20} U
          + q_{21} h_0 \label{eq:evol_V}\\
\nonumber \\
{\dot U} &=& s_{00} h_1
           + s_{01} h_1' \nonumber \\
         &+& s_{10} U
           + s_{11} h_0
           + s_{12} h_0'
           + s_{13} T_1 \nonumber \\
         &+& s_{20} V
           + s_{21} R
           + s_{22} H_1
           + s_{23} H_1'
           + s_{24} R'
           + s_{25} F_1
           + s_{26} S_1
           \label{eq:evol_U}
\end{eqnarray}
Here the various coefficients involved have the meaning described
earlier. In other words the ones of the form $a_{0i}$, $b_{0i}$,
$c_{0i}$, $d_{0i}$, $e_{0i}$, $k_{0i}$, $f_{0i}$, $q_{0i}$ and
$s_{0i}$ do not include rotational corrections. Those of the form
$a_{1i}$, $b_{1i}$, $c_{1i}$, $d_{1i}$, $e_{1i}$, $k_{1i}$,
$f_{1i}$, $q_{1i}$ and $s_{1i}$ include rotational corrections and
related to the perturbation functions which have the same parity
as the function that is evolved in every equation. Finally, those
of the form $a_{2i}$, $b_{2i}$, $c_{2i}$, $d_{2i}$, $e_{2i}$,
$k_{2i}$, $f_{2i}$, $q_{2i}$ and $s_{2i}$  include also rotational
corrections and correspond to the other parity and for spherical
harmonic indices $l \pm 1$. This last set of coefficients includes
the operators ${{\cal L}_1^\pm}$, ${{\cal L}_2^\pm}$ and ${{\cal
L}_3^\pm}$ which are responsible for the coupling with the $l \pm
1$ perturbations. All these coefficients are given explicitly in
Appendix II.

The above system of evolution equation is, as expected,
considerably more complicated than the one derived by Ruoff et
al.\cite{RSK02}, using the BCL gauge in the ADM form of Einstein's
equations. Still, it has considerable merits since parts of this
system have been tested and show to be stable. For example, in the
non-rotational case it has been used by Allen et
al.\cite{Allen98}. The part that corresponds to axial
perturbations (including rotations) have been used in earlier
papers by Ruoff and Kokkotas \cite{RK01,RK02} and Kokkotas et
al.\cite{KRA04}, while the evolution equations for the fluid have
been tested by Ruoff et al.\cite{RSK03} in the Cowling
approximation. Although, we don't use in this paper the above
system of equations in its full form, we will test, in the next
section, the part that deals with the evolution of spacetime
perturbations. This part is shown to be numerically unstable in
the BCL gauge, while in the present form it shows stability which
allows us to calculate the w-modes for rotating relativistic
stars.

\section{Inverse Cowling Approximation}

As we discussed earlier a first  and crucial test of the evolution
equations derived here will be the evolution of the spacetime
equations decoupled from the fluid ones. This simplification is
named {\em Inverse Cowling Approximation} (ICA) and it has been
suggested by Andersson et al.\cite{AKS96} as an easier way to
understand the spacetime perturbations and their characteristic
modes, the w-modes\cite{KS92}. It has been shown\cite{KS92,AKS96}
that the w-modes do not excite significant fluid motions and the
omission of the fluid perturbations is a legitimate approximation.

As discussed by Ruoff et al.\cite{RSK02} the various families of
fluid modes which are not degenerate in the non-rotating limit
i.e. the f, p and g-modes are not affected in a crucial way by the
coupling to the $l\pm1$ perturbations. This is not true for the
inertial (or rotational) modes, for those the coupling to the
various $l\pm1$ terms is vital for a proper understanding of the
spectrum, see also a detailed discussion by Lockitch and
Friedman\cite{lo99,LF99}.

The w-modes are similar to the quasi-normal modes of the black
holes where the rotation splits and shifts the spectra but (at
least in slow-rotation limit) the coupling with the various $l$'s
does not affect in a critical way the spectrum. Based on the above
arguments we studied the w-mode spectrum of rotating relativistic
stars.

The simplified version of the evolution equations for the
spacetime perturbations is:
\begin{eqnarray}
{\dot F}   &=& F_1 \qquad
{\dot F_2}  = F_1' \\
{\dot S}   &=& S_1 \qquad
{\dot S_2}  = S_1'\\
%
{\dot F_1} &=& e^{\nu-\lambda}F_2' + \frac{e^\nu}{r^3}\left(
3\beta+\alpha+4M-\Lambda r \right) F \nonumber \\
&+& {2\over r}\left(\alpha+\beta-re^{-\lambda}\right)S
+ {e^\nu \over r^2}(\alpha -\beta) F_2 \nonumber \\
           &+& {\I m\omega \over r}\left[(\alpha-\beta) H_1
           + r^2 e^{-\lambda} H_1' -r^3 e^{-\nu} S_1\right]
           \nonumber \\
           &+& { 4\I m \over r \Lambda}\left[(\alpha+\beta- r\Lambda
-re^{-\lambda})\omega -r^2 e^{-\lambda}\omega' \right] F_1
\label{eq:evol_F1_ica}
\\
%
{\dot S_1} &=& e^{\nu-\lambda}S_2'
            + {e^\nu \over r^3}\left(3\alpha +\beta -r \Lambda\right)S
\nonumber \\
            &+&{4 e^{2\nu}\over r^6}\left(e^\lambda \alpha^2-2Mr+r\beta \right) F
            + {e^\nu \over r^2}(\alpha-\beta) S_2
            \nonumber \\
           &-& \frac{2\I m}{r}e^{\nu-\lambda}\omega'H_1
           - \frac{2\I m}{r\Lambda}\omega
           \left[2(\alpha+\beta)+\Lambda r -2re^{-\lambda}\right]S_1
           \nonumber \\
           &+& \frac{4\I m e^\nu}{r^3 \Lambda}\left[2r^2e^{-\lambda}\omega' +
2\varpi(\alpha+\beta)+\omega(\alpha-\beta-2M)\right] F_1
\label{eq:evol_S1_ica}
\\
%
{\dot H_1} &=& \frac{2\alpha}{r^3}e^{\nu+\lambda} F + S + rS_2 \nonumber \\
           &-& \I m\omega H_1
           + \frac{\I m}{r\Lambda}\left[(2\alpha e^\lambda+r)\omega-r^2\omega'\right] F_1
           +  \frac{\I m}{r\Lambda} r^2\omega F_1'
\label{eq:evol_H1_ica}
\end{eqnarray}
where
\begin{equation} \Lambda=l(l+1)\ , \quad
\alpha \equiv M+kpr^3 \ \
\mbox{and} \ \ \beta \equiv-\left(M-k\rho r^3\right)
\end{equation}
Additionally, the following Hamiltonian constraint equation can be
used to set up the initial data for the evolution:
\begin{eqnarray}
F_2'&-&\frac{\beta}{r^2}e^{\lambda} F_2 + \frac{e^{\lambda}}{r^3}\left(3\beta+2M-r\Lambda\right)F \nonumber \\
&+&\frac{1}{2r}e^{\lambda-\nu}\left[4(\alpha+\beta)-4re^{-\lambda}-r\Lambda\right]S
-re^{-\nu}S_2 \nonumber \\
&+&
\frac{\I m}{r\Lambda}e^{\lambda-\nu}\left[4\omega(\alpha+\beta-re^{-\lambda})-3r\Lambda\omega-2r^2e^{-\lambda}\omega' \right]F1 \nonumber \\
&-&\I m \omega r^2 e^{\lambda-2\nu}S_1
+\frac{\I m}{2r}e^{-\nu}\left[2\omega(\alpha-\beta)e^{\lambda}-r^2\omega'\right]H_1
\nonumber \\
&-&\frac{2\I m}{\Lambda}e^{-\nu}r\omega F_1' +\I m \omega r e^{-\nu}H_1'=0
\end{eqnarray}
\begin{table}[htbp]
      \begin{center}
          \leavevmode
\begin{tabular}{|c|c|c|c|c|c|c|}
   \hline \hline
 $\epsilon$ & $\Omega$& $\sigma_{m=-2}$ & $\sigma_{m=-1}$ &$\sigma_{m=1}$ &$\sigma_{m=2}$ \\
 \hline
 0.0  &  0      & 10.8040  &  10.8040 & 10.8040 & 10.8040 \\
 0.2  &  0.4995 & 11.5294  &  11.1413 & 10.5070 & 10.2247 \\
 0.4  &  0.9902 & 12.5044  &  11.5294 & 10.2245 &  9.7312 \\
 0.6  &  1.4853 & 13.4466  &  11.9802 &  9.9763 &  9.3629 \\
 0.8  &  1.9804 & 14.2875  &  12.4622 &  9.7311 &  8.9827 \\
 1.0  &  2.4755 & 14.9495  &  12.9565 &  9.5126 &  8.7620 \\
 1.2  &  2.9706 & 15.5042  &  13.4471 &  9.3263 &  8.5791 \\
  \hline \hline
 \end{tabular}
          \caption{The frequencies of the $l=2$ polar w-mode for various rotation rates  for the polytropic model  $R$=8.86km and $M=1.267M_\odot$.
          The frequencies of the modes and the rotational velocity are given in kHz.}
          \label{table:m3}
      \end{center}
\end{table}


In our numerical study we have used polytropic equation of state
of the form $p=K\rho^\Gamma$ and uniform density stellar models
with typical masses and radii.  The numerical code was stable for
all values of the rotational parameter $\epsilon$.

In Fig.1 we present the evolution of the frequency of the first
w-mode as a function of the rotational parameter $\epsilon$. The
model that we have used has a polytropic equation of state with
$\Gamma=2$, $K=100$km$^{-2}$, radius $R=8.86$km and mass
$M=1.27M_\odot$. We observe the splitting of the frequency for
$m=\pm 1$ and $m=\pm 2$, while the frequency increases/decreases
linearly with respect to the rotational parameter $\epsilon$. In
the level of approximation that we have used the axisymmetric
modes ($m=0$) are not affected by rotation.

 In general the frequencies of rotating stars follow a
pattern of the form
\begin{equation}
\sigma=\sigma_0 -\kappa m\Omega
\end{equation}
where $\sigma_0$ is the frequency of the non-rotating star and
$\kappa=\kappa(M,R,...)$ is a function depending on the details of
the specific stellar model.


\begin{figure}[h]
\centering 
\epsfxsize=7cm 
\epsfysize=8cm 
\epsffile{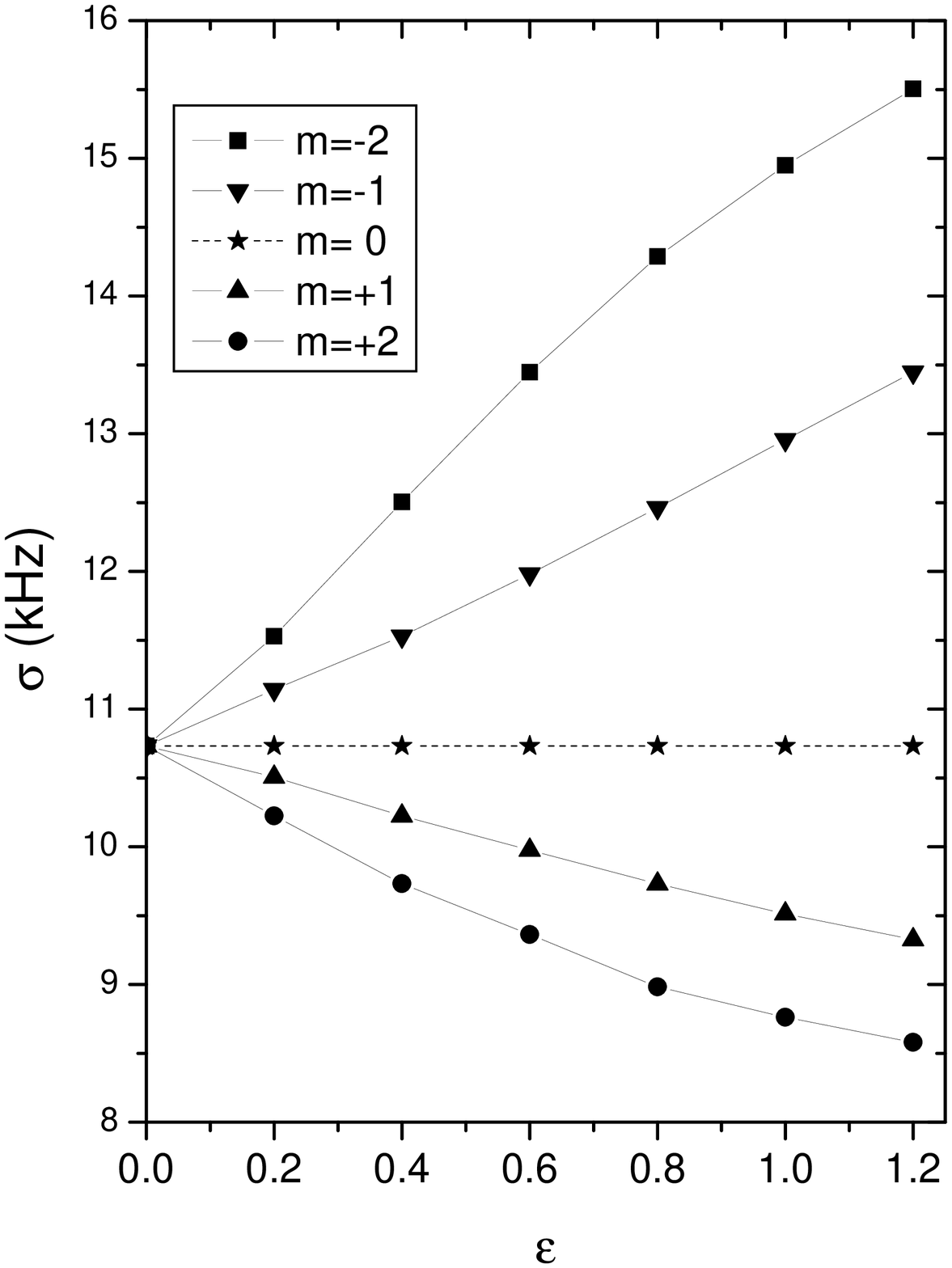}
\fcaption{The $l=2$ polar w-mode frequencies of a rotating polytropic star ($R=8.86$km
 and $M=1.267M_\odot$) as functions of the rotation parameter $\epsilon$.
}\label{fig_m3}
\end{figure}

In Fig.2 we show the frequency for two uniform density models.
Both have mass $M=1.4M_\odot$ one with $M/R=0.20$ and a more
compact one with $M/R=0.25$. One can easily observe that the more
compact the model the higher is the w-mode frequency $\sigma_0$ of
the non-rotating star as was shown by Kokkotas and Schutz\cite{KS92}.

The rotational corrections with $m>0$ for high rotation rates can
lead to negative frequencies of the modes which signal the onset
of an instability. It is apparent from Fig. 2 that the more
compact the model is the faster the frequency tends to zero i.e.
to the instability. For the models discussed here the instability
will never be excited for physically acceptable values of rotation
i.e. for $\epsilon <1$. Axial w-modes for the polytropic model
used here as well as for uniform density stars have been
calculated already\cite{Nils98,KRA04}. Actually, the ultra-compact
uniform density models, $M/R>0.4$, can become secularly unstable
as has been shown by Kokkotas et al.\cite{KRA04}.

\begin{figure}[h]
\centering
\epsfxsize=7cm
\epsfysize=8cm
\epsffile{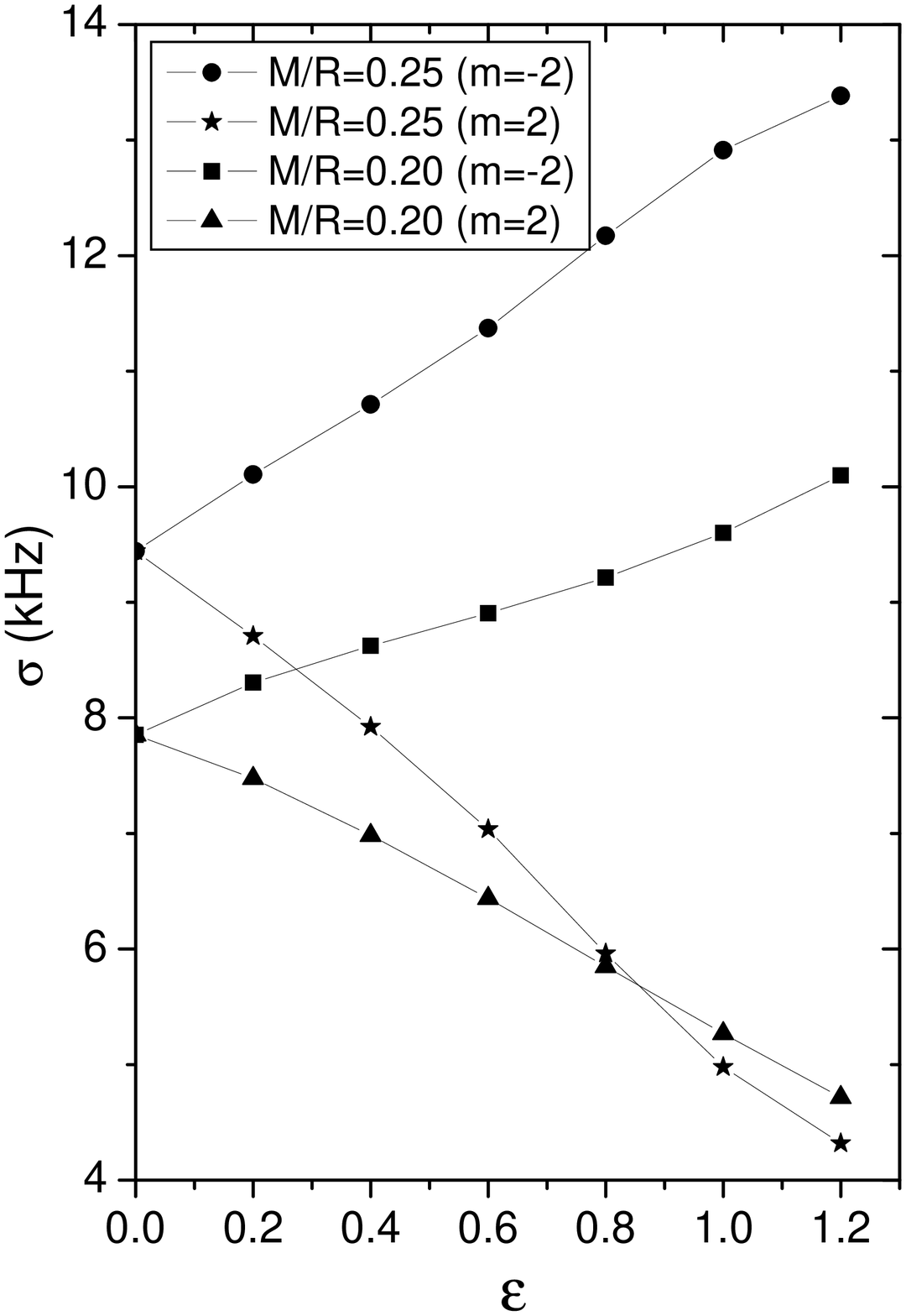}
\fcaption{The $l=2$ polar w-mode frequencies of two uniform density rotating stars 
($R=10.34km$ and $M=1.4M_\odot$) as functions of the rotation parameter $\epsilon$. Both models have mass $M=1.4 M_\odot$, the more comact model has radius  $R=8.27$km ($M/R=0.25$) and the other $R=10.34$km ($M/R=0.20$).}\label{fig_500}
\end{figure}

\section{Discussion}

In this article we have studied perturbation of slowly rotating
relativistic stars. In an attempt to find a stable system of
evolution equations we have derived a quite long but seemingly
stable hyperbolic system. In our calculations we have rederived
the linearized form of Einstein's equations for a slowly rotating
star.  This form is identical to the one by Kojima\cite{k92} and
for this purpose we refer the reader to his article. For the
evolution of the fluid perturbations we have derived the full set
of equations in Appendix I.

Finally, we have tested  a reduced version of the above system by
calculating the w-mode frequencies for slowly rotating stars in
the Inverse Cowling Approximation (ICA).

Concluding we have to admit that although the various parts of the
derived hyperbolic system of equations are well tested and are
stable, the whole system of equations involving the coupling with
higher and lower $l$'s is yet to be studied. This by itself is a
difficult task though the benefits suggest that it worth the
effort. An interesting question that has to be answered is whether
the instability observed in the ADM formalism of the equations in
the BCL gauge is due to the choice of the gauge or the fundamental
instability of the ADM formalism.

From the full system of equations one can study all types of modes
(f, p, w and inertial) including their rotational corrections and
their couplings. This is a very useful information in
understanding the behavior of the various modes as function of the
rotation of the star. It can actually contribute in the
gravitational asteroseismology\cite{astero1,astero2},
by predicting the angular velocity of the
star via the shifting of the various mode frequencies.


\nonumsection{Acknowledgements} \noindent We are grateful to V.
Ferrari, J. Ruoff,  U. Sperhake and N. Stergioulas
for insightful comments. This work has been supported by the EU
Programme 'Improving the Human Research Potential and the
Socio-Economic Knowledge Base' (Research Training Network Contract
HPRN-CT-2000-00137). KDK acknowledges support through the Center
of Gravitational Wave Physics, which is funded by the NSF number
cooperative agreement PHY 01-14375.

\nonumsection{References} \noindent

\def\prl#1#2#3{{ Phys. Rev. Lett.\ }, {\bf #1}, #2 (#3)}
\def\prd#1#2#3{{ Phys. Rev. D}, {\bf #1}, #2 (#3)}
\def\plb#1#2#3{{ Phys. Lett. B}, {\bf #1}, #2 (#3)}
\def\prep#1#2#3{{ Phys. Reports}, {\bf #1}, #2 (#3)}
\def\phys#1#2#3{{ Physica}, {\bf #1}, #2 (#3)}
\def\jcp#1#2#3{{ J. Comput. Phys.}, {\bf #1}, #2 (#3)}
\def\jmp#1#2#3{{ J. Math. Phys.}, {\bf #1}, #2 (#3)}
\def\cpr#1#2#3{{ Computer Phys. Rept.}, {\bf #1}, #2 (#3)}
\def\cqg#1#2#3{{ Class. Quantum Grav.}, {\bf #1}, #2 (#3)}
\def\cma#1#2#3{{ Computers Math. Applic.}, {\bf #1}, #2 (#3)}
\def\mc#1#2#3{{ Math. Compt.}, {\bf #1}, #2 (#3)}
\def\apj#1#2#3{{ Astrophys. J.}, {\bf #1}, #2 (#3)}
\def\apjl#1#2#3{{ Astrophys. J. Lett.}, {\bf #1}, #2 (#3)}
\def\apjs#1#2#3{{ Astrophys. J. Suppl.}, {\bf #1}, #2 (#3)}
\def\acta#1#2#3{{ Acta Astronomica}, {\bf #1}, #2 (#3)}
\def\sa#1#2#3{{ Sov. Astro.}, {\bf #1}, #2 (#3)}
\def\sia#1#2#3{{ SIAM J. Sci. Statist. Comput.}, {\bf #1}, #2 (#3)}
\def\aa#1#2#3{{ Astron. Astrophys.}, {\bf #1}, #2 (#3)}
\def\apss#1#2#3{{Astrop. Sp. Sci.}, {\bf #1}, #2 (#3)}
\def\mnras#1#2#3{{ Mon. Not. R. Astr. Soc.}, {\bf #1}, #2 (#3)}
\def\prsla#1#2#3{{ Proc. R. Soc. London, Ser. A}, {\bf #1}, #2 (#3)}
\def\ijmpc#1#2#3{{ I.J.M.P.} C {\bf #1}, #2 (#3)}
\def\novcb#1#2#3{{ Nuovo Cim. B} {\bf #1}, #2 (#3)}
\def\lrr#1#2#3{{ Living Rev. in Rel.} {\bf #1}, #2 (#3)}
\def\ptp#1#2#3{{ Progress Theor. Phys.} {\bf #1}, #2 (#3)}

\newpage

\section{Appendix I: Equations for fluid perturbations }

In this Appendix we derive the perturbations of equations of
motion for the fluid. The will be derived from the variation of
the conservations laws of the energy-momentum tensor, i.e. from
equation (\ref{eq:pert_fluid}):
\begin{equation}
\delta \left( {T^{\mu \nu}}_{;\nu} \right)=0
\end{equation}
The variation of the equations describing the conservation of
energy and the radial component of the momentum will be written
as:
\begin{equation}
A_{lm}^{(I)}Y_{lm} + B_{lm}^{(I)} \sin(\theta)
\partial_{\theta} Y_{lm} + C_{lm}^{(I)}\partial_\phi Y_{lm}=0
\label{eq:fluid_1_2}
\end{equation}
where $I=4,5$.
We use these values for $I$ because the above two
equations can have the same form as equation (14) in
Kojima\cite{k92} and can be considered as a complementary set to
these equations.
Next we shall decompose the above equations to a
specific mode with $l$, $m$.
Multiplying equation
(\ref{eq:fluid_1_2} with $Y_{lm}^*$ and integrating over the solid
angle, we get:
\begin{equation}
A_{lm}^{(I)}Y_{lm} + \I m C_{lm}^{(I)} \sin(\theta) + Q_{l-1
m}(l-1)B^{(I)}_{l-1 m}-Q_{l+1 m}(l+2)B^{(I)}_{l+1 m} =0
\end{equation}
which has similar structure as equation (20) of
Kojima\cite{k92}. In our convention this equation will be written
as:
\begin{equation}
A_{lm}^{(I)}Y_{lm} + \I m C_{lm}^{(I)} \sin(\theta) + {{\cal
L}_1^{\pm}}B^{(I)}_{lm} =0
\end{equation}

The variations of the fluid equations for the $(\theta)$ and
$(\phi)$ components have the form:
\begin{eqnarray}
\left( \alpha_{lm}^{(J)} + {\tilde
\alpha}_{lm}^{(J)}\cos(\theta)\right)\partial_\theta Y_{lm}
 &-&\left(\beta_{lm}^{(J)}+{\tilde \beta}_{lm}^{(J)}\cos(\theta)\right)
 \sin^{-1}\theta \partial Y_{lm} \nonumber \\
 &+& \eta_{lm}^{(J)}\sin \theta
 Y_{lm} + \xi_{lm}^{J} X_{lm}
\nonumber \\
 &+& \chi_{lm}^{(J)} \sin(\theta)
 W_{lm}=0 \\
\left( \beta_{lm}^{(J)} + {\tilde
\beta}_{lm}^{(J)}\cos(\theta)\right)\partial_\theta Y_{lm}
 &+&\left(\alpha_{lm}^{(J)}+{\tilde \alpha}_{lm}^{(J)}\cos(\theta)\right)
 \sin^{-1}\theta \partial Y_{lm} \nonumber \\
 &+& \zeta_{lm}^{(J)}\sin \theta
 Y_{lm} + \chi_{lm}^{J} X_{lm}
\nonumber \\
  &-& \xi_{lm}^{(J)} \sin(\theta)
 W_{lm}=0
\end{eqnarray}
which have the structure of equations (15) and (16) of
Kojima\cite{k92}, therefore we will index them as $J=2$. Their
decomposition into spherical harmonics leads to equations of the
form
\begin{eqnarray}
\Lambda \alpha_{lm}^{(J)}&+& \I
m\left[(\Lambda-2)\xi_{lm}^{(J)}-{\tilde \beta}_{lm}^{(J)}
-\zeta_{lm}^{(J)}\right] \nonumber \\
&+& {\cal L}_2^{\pm} \left[\eta_{lm}^{(J)}-(\Lambda^\pm-2)
\chi_{lm}^{(J)}\right] +{\cal L}_3^{\pm} {\tilde
\alpha}_{lm}^{(J)} =0
\\
\Lambda \beta_{lm}^{(J)}&+&\I
m\left[(\Lambda-2)\chi_{lm}^{(J)}+{\tilde
\alpha}_{lm}^{(J)} +\eta_{lm}^{(J)}\right] \nonumber \\
&+& {\cal
L}_2^{\pm}\left[\zeta_{lm}^{(J)}+(\Lambda^\pm-2)\xi_{lm}^{(J)}
\right]+ {\cal L}_3^{\pm} {\tilde \beta}_{lm}^{(J)}=0
\end{eqnarray}
where $\Lambda=l(l+1)$ and $\Lambda^\pm$ stands for $\Lambda$ where $l\rightarrow l+1$ or $l\rightarrow l-1$ i.e. $\Lambda^+=(l+1)(l+2)$ and $\Lambda^-=(l-1)l$.  
The above set of equations has the same structure as equations (24)-(25) in \cite{k92}.

The various functionals used in the above sets of equations have
the following form:
\begin{eqnarray}
A_{lm}^{(4)} &=& -{\dot \rho_1} + 2 \rho {\dot H_0}
-\frac{1}{k}e^{\nu-\lambda} R' +e^{-\lambda}(p-\rho) H_1'+
             \frac{e^{\nu}\Lambda}{kr^2}V \nonumber \\
	     &-& (\rho+p)\left( {\dot K} +\frac{1}{2}{\dot H_2}\right)
	     -\frac{e^{\nu-\lambda}}{2kr}\left(3r\nu'-r\lambda'+4\right)R
	     \nonumber \\
             &-& \frac{e^{-\lambda}}{2r}\left[r\nu'(p+\rho)C_s^{-2} -
             2r\nu'\rho +(\rho-p)(r\lambda'-4)\right]H_1 \nonumber \\
                          \\
B_{lm}^{(4)} &=& 
\frac{e^{-\lambda}}{2r}\left\{(\rho+p)\left[\left(\Omega(r\lambda'+rC_s^{-2}\nu'-4\right)-2r\omega(\lambda'+\nu')\right] +8\omega p\right\} h_1 \nonumber \\
&+& 4e^{-\nu}\omega \rho {\dot h_0}
               - \frac{2}{k}\varpi{\dot U}
               +  e^{-\lambda}\left[2\omega p-(\rho+p)\Omega \right]h_1'
\\
C_{lm}^{(4)} &=& -\frac{2}{k}\varpi{\dot V}
   +\left[\frac{3}{2}(\rho+p)\varpi+2\rho\omega \right]H_0 -\rho_1\Omega -p_1 \varpi 
   \nonumber \\
   &-& 2\left(\Omega \rho +\varpi p \right)K
  \\
\nonumber \\
 A_{lm}^{(5)} &=& \frac{1}{k}\dot{R} + 2e^{-\nu}\rho {\dot
 H_1}+ 2pH_2' -\frac{4}{r}p K -\nu'\rho H_0 +p_1'
\nonumber \\
 &&+\frac{\nu'}{2}\left(\rho_1+p_1\right) 
 + \left(\frac{4}{r}p-\rho\nu'\right) H_2-\frac{1}{2}\left(\rho+p\right)H_0'\\
B_{lm}^{(5)} &=& 2e^{-\nu}\left(\rho \Omega + p\varpi\right) {\dot
h_1} + \frac{1}{k r}\left(r\varpi \nu'+r\omega'-2\varpi\right)U \nonumber \\
&-& e^{-\nu}\left(\rho+p\right)\Omega h_0' \nonumber \\
&+&
 \frac{e^{-\nu}}{r}\left[(\rho-p)(\omega'-\omega \nu')  -\frac{2}{r}( \rho+p) \varpi\right]h_0\\
C_{lm}^{(5)} &=& \frac{1}{k}\Omega R 
+ 2e^{-\nu}\left(\Omega \rho +\varpi p \right)H_1 + \frac{1}{kr}\left(r\nu'\varpi+r\omega'-2\varpi\right)V
\end{eqnarray}

\begin{eqnarray}
\alpha_{lm}^{(2)}&=& {\dot V} +k p_1 + 2 k p K-\frac{1}{2}k(\rho+p)H_0 \\
{\tilde a} _{lm}^{(2)}&=& k e^{-\nu} \left[2\omega \rho -(\rho+p)\Omega \right] h_0  +(2\omega-\Omega )U  \\
\beta_{lm}^{(2)} &=& {\dot U} + 2k e^{-\nu} \rho {\dot h_0} + 2k p
e^{-\lambda} h_1' - k e^{-\lambda} \left(p\lambda' + \rho \nu' - \frac{4}{r} p \right) h_1 \\
{\tilde \beta}_{lm}^{(2)} &=& (2\varpi-\Omega)V \\
\eta_{lm}^{(2)}&=& k e^{-\nu}\Lambda\left(\frac{3}{2}(\rho+p) \Omega-p\omega\right)h_0 +\frac{1}{2}\Lambda \Omega U \\
\xi_{lm}^{(2)}&=& \frac{1}{2}\Omega V \\
\chi_{lm}^{(2)}&=& k e^{-\nu}\left(\frac{1}{2}(\rho+p) \Omega-p\omega\right)h_0 +\frac{1}{2} \Omega U \\
\zeta_{lm}^{(2)}&=&k r^2 e^{-\lambda-\nu} 
{\mbox {\Large \{} }
(\rho+p)\left[\varpi' - \left(\nu'-\frac{4}{r}\right)\varpi - \frac{1}{2}\left(C_s^{-2}\nu'-\lambda'\right)\Omega\right] 
\nonumber  \\
&+& 
\omega\left(p\lambda'+\rho\nu'+\frac{2}{r}(\rho-p)\right) 
{\mbox {\Large \}}}
H_1 \nonumber \\
&+&\frac{1}{2}re^{-\lambda}\left[4(\Omega+\varpi)-2r\omega'+\left(2\omega+\Omega\right)r\nu'-\lambda'r\Omega\right]W \nonumber \\
&-& \frac{1}{2}k r^2 e^{-\nu}\left[(\rho+p)\varpi -4\rho \omega\right]{\dot H_0}
+k r^2 e^{-\nu}\left(\varpi{\dot p_1} +\Omega {\dot \rho_1}\right)
\nonumber \\
&+& k r^2 e^{-\nu}\left[3(p+\rho)\Omega-2p\omega\right]{\dot K}
\nonumber \\
&+& k r^2 e^{-\lambda-\nu}\left[(\rho+p)\varpi+(\rho-p)\omega\right]H_1'
\ .
\end{eqnarray}

Some further simplifications are still possible in the above two
relations. Therefore the terms related to the rotational
corrections get the form:
\begin{eqnarray}
{\cal L}_3^{\pm} {\tilde \alpha}_{lm}^{(J)} &+& {\cal L}_2^{\pm}
\left[\eta_{lm}^{(J)}-(\Lambda^\pm-2)
\chi_{lm}^{(J)}\right] \nonumber \\
&=& ke^{-\nu}\left\{\left[\Omega( \rho-p)-2\varpi \rho\right]{{\cal
L}_3^{\pm}}
+ \left[(\rho+p)(\Lambda^\pm+1)\Omega-2p\omega\right]{{\cal L}_2^{\pm}}\right\} h_0 \nonumber \\
&+&\left[(\Omega-2\varpi){{\cal L}_3^{\pm}}+\Omega {{\cal
L}_2^{\pm}}\right]U
\end{eqnarray}
and
\begin{eqnarray}
 (\Lambda-2)\chi_{lm}^{(J)}+{\tilde \alpha}_{lm}^{(J)}
+\eta_{lm}^{(J)} &=&
 k e^{-\nu}\left[2\rho\omega-p\Lambda \omega+\frac{1}{2}(3\Lambda-2)(\rho+p)\right]h_0 
\nonumber \\
 &+&\frac{1}{2}(\Lambda-2)\Omega V +\frac{1}{2}\left[4\omega+(\Lambda-2)\Omega\right]U \ .
\end{eqnarray}

In section 2 the above set of equations is simplified still
further and we will use instead of the functions $K_{lm}$ and
$H_{0,lm}$ the new ones i.e. $F_{lm}$ and $S_{lm}$.

\newpage

\section{Appendix II: Coefficients of the evolution equations}
In this appendix we provide the background coefficients for the
evolution equations (\ref{eq:evol_1})- (\ref{eq:evol_U}).

We have used the following substitutions for the pressure and
density
\begin{equation}
\alpha \equiv M+kpr^3=\frac{1}{2}r^2 e^{-\lambda}\nu' \ \
\mbox{and} \ \ \beta \equiv-\left(M-k\rho r^3\right)=\frac{1}{2}r^2
e^{-\lambda}\lambda'
\end{equation}
which simplify considerably some of the equations.

Background coefficients for the evolution equation
(\ref{eq:evol_F1}) for $F_1$
\begin{eqnarray}
a_{01}&=&\frac{e^\nu}{r^3}\left[ 3\beta+\alpha+4M-\Lambda r \right] \\
a_{02}&=& {2\over r}\left(\alpha+\beta-re^{-\lambda}\right)\\
a_{03}&=& \frac{2}{r^2}e^{\nu} (\alpha+\beta)(C_s^{-2}-1))\\
a_{04}&=& {e^\nu \over r^2}(\alpha -\beta) \\
a_{10}&=& {8\I m\over \Lambda}e^{\nu-\lambda}r^2 \varpi \\
a_{11}&=& 4\I m r e^{\nu}\omega + {16\I m\over
\Lambda}e^{\nu}\varpi
\left(\beta-re^{-\lambda}\right) -\frac{8\I m}{\Lambda}e^{\nu-\lambda}r^2\varpi' \\
a_{12}&=& {\I m\over r}(\alpha-\beta)\omega \\
a_{13}&=&-{8\I m\over \Lambda}e^{\nu-\lambda}r^2 \varpi :=-a_{10}\\
a_{14}&=& i m r \omega e^{-\lambda}\\
a_{15}&=& { 4\I m \over r \Lambda}\left[(\alpha+\beta- r\Lambda
-re^{-\lambda})\omega -r^2 e^{-\lambda}\omega' \right]\\
a_{16}&=&-\I mr^2 e^{-\nu}\omega \\
%
a_{20}&=&
\frac{4}{r^2}\left[\alpha\Omega +\beta\varpi\right]{{\cal
L}_1^\pm} + \frac{2}{\Lambda
r^2}\left(3r^2e^{-\lambda}\omega'-2\alpha\omega\right){{\cal
L}_3^\pm} \nonumber \\
&+&\frac{2}{\Lambda r}e^{-\lambda}\omega'
\left[8re^{-\lambda}+4(2\alpha+3\beta)+r(\Lambda^\pm-\Lambda-1)\right]
{{\cal L}_2^\pm} \nonumber \\
&+& \frac{4}{\Lambda r^3}(\alpha+\beta) \left[2\alpha
C_s^{-2}-\Lambda r + 2(3\alpha+2\beta)\right]\varpi{{\cal L}_2^\pm}
\nonumber \\
&+& \frac{4}{\Lambda r^2}\left[\Lambda^\pm(re^{-\lambda}-\beta)-\alpha \right]\omega{{\cal L}_2^\pm}
\\
a_{21} &=& {\omega e^{\nu-\lambda}\over r^2}(\alpha-\beta)
\left(2 {{\cal L}_1^\pm}+3\frac{\Lambda^\pm}{\Lambda}{{\cal L}_2^\pm}\right)\\
a_{22} &=& 4re^{\nu}\varpi {{\cal L}_1^\pm} + {4 e^{\nu}\over
\Lambda}
\left[(4\beta-4re^{-\lambda}-r\Lambda)\varpi
+ 4e^{-\lambda} r^2 \omega' \right]{{\cal L}_2^\pm} \\
a_{23}
 &=&-\frac{e^{-\lambda}}{r\Lambda}\left[2r(1-\Lambda^\pm)\omega
 -16\varpi(\alpha+\beta)
- (8re^{-\lambda}-r\Lambda +4 \beta)r\omega'\right]  {{\cal L}_2^\pm}
\nonumber \\
&-& 2e^{-\lambda}(\omega+r\omega') {{\cal L}_1^\pm}
+ \frac{2}{\Lambda}e^{-\lambda}\omega {{\cal L}_3^\pm}\\
a_{24} &=& \omega  e^{\nu-2\lambda} \left(2 {{\cal
L}_1^\pm}+ 3\frac{\Lambda^\pm}{\Lambda}{{\cal L}_2^\pm}\right) \\
a_{25} &=& -8e^{\nu-\lambda}r^2\varpi {{{\cal L}_2^\pm}\over \Lambda}\\
a_{26} &=& re^{-\lambda}\omega' {{\cal L}_1^\pm}
+ \frac{2}{\Lambda}e^{-\lambda}\omega {{\cal L}_3^\pm}\nonumber \\
&+&{e^{-\lambda}\over r\Lambda} \left[ 8(\alpha+\beta)\varpi
+r(8re^{-\lambda}-r\Lambda +4\beta)\omega'
-r(\Lambda^\pm-2)\omega\right]{{\cal L}_2^\pm}
\end{eqnarray}

Background coefficients for the evolution equation
(\ref{eq:evol_S1}) for $S_1$
\begin{eqnarray}
b_{01}&=& {e^\nu \over r^3}\left(3\alpha +\beta -r \Lambda\right) \\
b_{02}&=& {4 e^{2\nu}\over r^6}\left(e^\lambda \alpha^2-2Mr+r\beta \right)\\
b_{03}&=& {e^\nu \over r^2}(\alpha-\beta)\\
b_{10}&=&-\frac{4\I
m}{r\Lambda}e^{2\nu}(\omega+2\varpi)(3\alpha-\beta+2re^{-\lambda})C_s^2
  +\frac{8\I m}{\Lambda}re^{2\nu-\lambda}\varpi' \nonumber \\
 &+&\frac{4\I m}{r\Lambda}e^{2\nu}\left[2\varpi(2\alpha+re^{-\lambda}-\beta)
 +(3\alpha-\beta+2re^{-\lambda})\omega\right]\\
b_{11}&=& {8\I m\over r^3 \Lambda}e^{2\nu}\left[
\left(2\alpha^2e^\lambda+6r\alpha-\Lambda r^2\right)\varpi
      + r^2\left(\alpha+re^{-\lambda}\right)\varpi'\right]
      \nonumber \\
   &+&\frac{4\I m}{r}e^{2\nu}\left[(C_s^2-1)\omega+2\varpi C_s^2\right] \\
b_{12}&=&\frac{4\I m}{r^4\Lambda}e^{\nu}C_s^2(2\varpi+\omega)\left[(\alpha-\beta-2M)\left(2re^{-\lambda}-\beta\right)-2\alpha(\beta+M)\right]
\nonumber \\
&+&\frac{4\I m}{r^4 \Lambda}e^{\nu}\alpha(\alpha+\beta)\left(2\varpi-C_s^{-2}\omega\right) -\frac{2\I m}{r}e^{\nu-\lambda}\omega'
\nonumber \\
&-&\frac{4\I m}{r^4\Lambda}e^{\nu}\left[2(\alpha-\beta)(r-3M-2\beta)-\alpha^2-3\beta^2-4Mre^{-\lambda}\right]\omega
\\
b_{13}&=& {4\I m\over \Lambda}e^{2\nu-\lambda}r(\omega+2\varpi)(1-C_s^2) \\
b_{14}&=& {8\I m e^{2\nu}\over r\Lambda}\varpi \left(re^{-\lambda}+\alpha\right)\\
b_{15}&=&\frac{4\I
m}{r^2\Lambda}e^{\nu-\lambda}(\alpha-2M-\beta)\left[2\varpi C_s^2
+ (C_s^2-1)\omega\right] \\
b_{16}&=&  \frac{8 \I m}{r \Lambda}e^{\nu-\lambda}\omega'
+\frac{2\I m}{r^3\Lambda}e^{\nu}(2\varpi+\omega)(4M+\beta-3\alpha)
\nonumber \\
&+&\frac{2 \I m}{r^3\Lambda}e^{\nu}\left[2(\beta+5\alpha-4M)\varpi
-(3\beta-5\alpha+8M)\omega\right]
\\
b_{17}&=& {2\I m\over r\Lambda}(2\varpi+\omega)(4M+3\beta-\alpha)C_s^2
-\frac{4\I m}{r\Lambda}(\alpha+\beta)\varpi
\nonumber \\
&-&\frac{2 \I m}{r\Lambda}\left(r\Lambda+\alpha+5\beta+8M-2r\right)\omega
\\
b_{20}&=&\frac{e^{\nu}}{r^5}\left\{ 4r(\alpha+\beta)\varpi +
2r^2(\alpha+2re^{-\lambda})\omega'-2\left[2\alpha(r+\alpha
e^\lambda) +r^2\Lambda\right]\omega\right\}{{\cal
L}_1^\pm}
\nonumber \\
&+&\frac{2e^\nu}{r^4\Lambda}\omega'\left[4\alpha^2+r(1-\Lambda)\alpha-2r^2\Lambda
e^{-\lambda}-3r^2e^{-\lambda}\left(1+4e^{-\lambda}-2\Lambda^\pm
\right)\right]{{\cal L}_2^\pm} \nonumber \\
&+& \frac{2 e^{\nu}}{r^4 \Lambda}\left[ \left(r\Lambda
+\frac{2}{r}\alpha^2e^{\lambda}+2\alpha \right)\omega -12
(\alpha+\beta)\varpi +r(\alpha-15re^{-\lambda})\omega' \right]
{{\cal L}_3^\pm}
\nonumber \\
&+&\frac{4e^{\nu}}{r^5\Lambda}(\alpha+\beta)\left[2\alpha-6r+8M+(\Lambda-2\Lambda^\pm)r\right]\varpi {{\cal L}_2^\pm}
\nonumber \\
&-&\frac{2e^{\nu}}{r^5\Lambda}\left[(2\alpha^2e^{\lambda}+r^2\Lambda+2r\alpha)(\Lambda^\pm-1)+2r\Lambda^\pm(r-4M+2\alpha+2\beta)\right]\omega {{\cal L}_2^\pm}
\nonumber \\
&-&\frac{8}{r^6\Lambda}e^{\nu+\lambda}\alpha(\alpha+\beta)\left[\alpha+(\alpha+re^{\lambda})C_S^{-2}\right]\varpi {{\cal L}_2^\pm}
\\
b_{21}&=& \frac{3
e^{(2\nu-\lambda)}}{r^3}\frac{\Lambda^\pm}{\Lambda}
(\alpha-\beta)\omega' {{\cal L}_2^\pm} 
+ \frac{3 e^{2\nu}}{r^5}\frac{\Lambda^\pm}{\Lambda}\alpha(\alpha+\beta)C_s^{-2}\omega{{\cal L}_2^\pm}
\nonumber \\
&+& \frac{3 e^{2\nu}}{r^5}\frac{\Lambda^\pm}{\Lambda}\left[r^2 e^{-\lambda}(\Lambda^\pm-\Lambda)-4rMe^{-\lambda} + 2\beta(r-\alpha-2M)\right]\omega {{\cal L}_2^\pm}
\\
b_{22}&=&-\frac{4}{r}e^{2\nu}\left[(\varpi+2\omega){{\cal
L}_1^\pm}- \frac{2}{\Lambda}\omega{{\cal L}_3^\pm}\right]
\nonumber \\
&+&\frac{4 e^{2\nu}}{\Lambda r^3} \left[2(1-2\Lambda^\pm)r^2\omega +
(4 e^{\lambda}\alpha^2+12 r\alpha -r^2\Lambda)\varpi)\right]{{\cal
L}_2^\pm}
\\
b_{23}&=&\frac{e^{\nu-\lambda}}{r^3}\left[2(r+\alpha
e^{\lambda})\omega -r^2\omega'\right]{{\cal L}_1^\pm}
+\frac{8}{r\Lambda}e^{\nu-\lambda}\omega'{{\cal L}_3^\pm}
\nonumber \\ &+&\frac{16
e^{\nu}}{r^4\Lambda}(\alpha+\beta)(\alpha+re^{-\lambda})\varpi
{{\cal L}_2^\pm} \nonumber \\
&-&\frac{2 e^{\nu}}{r^3
\Lambda}\omega\left[\alpha+re^{-\lambda}+\Lambda^\pm(\alpha-3re^{-\lambda})\right]{{\cal
L}_2^\pm} \nonumber \\
&+& \frac{e^{\nu-\lambda}}{r^3 \Lambda}\left[ r^2(\Lambda
+4-6\Lambda^\pm+12e^{-\lambda})-4\alpha^2 e^\lambda
\right]\omega'{{\cal L}_2^\pm}
\\
b_{24}&=& \frac{3}{r^3}\frac{\Lambda^\pm}{\Lambda}e^{2\nu-\lambda}\left[r^2e^{-\lambda}\omega'
+(3\alpha-4\beta-r e^{-\lambda})\omega \right]{{\cal L}_2^\pm}
\\
b_{25}&=&\frac{8e^{2\nu}}{r\Lambda}\left(\alpha+re^{-\lambda}\right)\varpi{{\cal
L}_2^\pm}
\\
b_{26}&=& \frac{e^\nu}{r^3}\left[2(\alpha-3re^{-\lambda})\omega-r^2
e^{-\lambda}\omega' \right]  {{\cal L}_1^\pm}
\nonumber \\
&+& \frac{2 e^\nu }{r^3\Lambda}
      \left[(re^{-\lambda}-\alpha-2\beta)\omega
       - r^2 e^{-\lambda}\omega' \right] {{\cal L}_3^\pm}
       \nonumber \\
       &+& \frac{e^\nu}{r^3\Lambda}\omega \left[
       (6\alpha+3\beta-7re^{-\lambda})\Lambda^\pm -2 (\alpha+\beta-r
       e^{-\lambda})\right]{{\cal L}_2^\pm}
       \nonumber \\
       &+& \frac{e^{\nu-\lambda}}{r^3\Lambda}\omega'\left[4\alpha^2 e^\lambda
       +r^2(\Lambda-\Lambda^\pm-2-12 e^{-\lambda} \right]{{\cal L}_2^\pm}
       \nonumber \\
      &+& \frac{8 e^{\nu}}{r^4\Lambda}\varpi(\alpha+\beta)
       \left[ (\alpha-\beta)e^{2\nu}-2\alpha +\beta - r e^{-\lambda}\right]{{\cal L}_2^\pm}
\\
b_{27}&=& \frac{e^{\nu-\lambda}}{r^2\Lambda}\left[
8(\alpha+\beta)(e^{2\nu}-1)\varpi + (4-3\Lambda^\pm)r\omega
\right]{{\cal L}_2^\pm} \nonumber \\
&+& \frac{4 e^{\nu-\lambda}}{r\Lambda}\omega {{\cal L}_3^\pm}
\\
b_{28}&=& \frac{3\omega}{r\Lambda}e^{2(\nu-\lambda)}\omega
\Lambda^\pm {{\cal L}_2^\pm}
\end{eqnarray}

Background coefficients for the evolution equation
(\ref{eq:evol_H1}) for $H_1$
\begin{eqnarray}
c_{00}&=& {2 \alpha \over r^3}e^{\nu+\lambda} \\
c_{01}&=&1 \\
c_{02}&=& r \\
c_{10} &=& -i m \omega \\
c_{11} &=& {8 i m \over \Lambda}
e^{\nu+\lambda}(re^{-\lambda}+\alpha)\varpi \\
c_{12} &=& {4\I m \over \Lambda}e^\nu r^2 \varpi \\
c_{13} &=& {\I m\over r\Lambda} \left[ (2\alpha e^\lambda
+r)\omega
-r^2\omega' \right] \\
c_{14} &=& {\I m \over \Lambda} r \omega \\
c_{20} &=& \frac{2}{\Lambda r^3}e^{\lambda}
\left[4(\alpha+\beta)(\alpha +re^{-\lambda})\varpi -
r\alpha\omega\right]{{\cal L}_2^\pm} \nonumber \\
&+& \frac{1}{r\Lambda}\left(4re^{-\lambda}+r(1-\Lambda^\pm)
+4\alpha \right)\omega'{{\cal L}_2^\pm}\nonumber
\nonumber \\
&+& \frac{1}{\Lambda r^2}\left(3r^2\omega'-2\alpha e^\lambda
\omega
\right){{\cal L}_3^\pm} \\
c_{21} &=& {8\varpi \over \Lambda}
e^{\nu+\lambda}(re^{-\lambda}+\alpha){{\cal L}_2^\pm} \\
c_{22}&=&{1\over \Lambda} \left[ (1-2\Lambda^\pm)\omega
-2(\alpha+re^{-\lambda})\omega' \right]{{\cal
L}_2^\pm} +\frac{1}{\Lambda}\omega {{\cal L}_3^\pm} \\
c_{23}&=&{1\over 2\Lambda} \left[ (2-\Lambda^\pm)\omega
+4(\alpha+re^{-\lambda})\omega' \right]{{\cal
L}_2^\pm} +\frac{1}{\Lambda}\omega {{\cal L}_3^\pm} \\
 c_{24}&=& {3\Lambda^\pm \over 2 \Lambda}e^{\lambda-\nu}\omega{{\cal
L}_2^\pm} \\
c_{25}&=& {3\Lambda^\pm \over 2 \Lambda r^2}e^{\nu}\omega(\alpha-\beta){{\cal
L}_2^\pm} 
\end{eqnarray}

Background coefficients for the evolution equation
(\ref{eq:evol_T1}) for $T_1$
\begin{eqnarray}
d_{01}&=& {e^{\nu+\lambda} \over
r^4}\left[(\alpha-3\beta)(a-re^{-\lambda})
+C_s^{-2}\alpha(\alpha+\beta) \right] \nonumber \\
&+&   {e^{\nu} \over r^2}(2e^{-\lambda}-\Lambda) \qquad \\
d_{02}&=& -{e^\nu \over
r^2}\left[2re^{-\lambda}+3(\beta-\alpha)\right]\\
d_{10}&=&-{2\I m\over
r\Lambda}\left[r\Lambda+2\alpha+4\beta+6re^{-\lambda}\right]\omega'
\nonumber \\
&-&{4\I m\over r^3\Lambda}(\alpha+\beta)e^{\lambda}
\left[C_s^{-2}\alpha +(\alpha+2re^{-\lambda})\right]\varpi \\
d_{11}&=&{3\I m\over 2r^2}e^{\nu}\omega(\beta-\alpha) \\
d_{12}&=&{8\I m\over \Lambda}e^{\nu}\left(\alpha
e^{\lambda}\varpi
-r^2 \omega'\right)\\
d_{13}&=&{2\I m \over
r\Lambda}\left[4(\alpha+\beta)\varpi-r(\alpha-3re^{-\lambda})\omega'\right]\\
d_{14} &=& -{3 \over 2} \I m e^{\nu-\lambda} \omega\\
d_{15}&=&{4\I m \over \Lambda}r^2\varpi e^{\nu}\\
d_{16}&=&{2\I m\over
r\Lambda}\left[r(\alpha-3re^{-\lambda})\omega'-2(\alpha+\beta)\varpi
\right]
-{1\over 2}\I m\omega\\
d_{20}&=&-{1 \over \Lambda} \omega
\left({{\cal L}_2^\pm}+{{\cal L}_3^\pm}\right) \\
d_{21}&=& -{4\over \Lambda}e^{\nu}\left(2e^{\lambda}\alpha
\varpi+r^2\varpi'\right) {{\cal L}_2^\pm}\\
d_{22}&=& {4\over \Lambda}e^{\nu}r^2 \varpi {{\cal L}_2^\pm} \\
d_{23} &=& -d_{22} \\
d_{24}&=&-{2\over \Lambda}r\omega' {{\cal L}_2^\pm}
\end{eqnarray}

Background coefficients for the evolution equation
(\ref{eq:evol_h0}) for $h_0$
\begin{eqnarray}
e_{00}&=&{e^\nu \over r^2}(\alpha-\beta) \\
e_{01}&=& e^{\nu -\lambda}\\
e_{10}&=&-\I m\omega +{\I m\over
r\Lambda}\left[4(\alpha+\beta)\varpi +
2r^2e^{-\lambda}\omega' -r\Lambda \omega\right] \\
e_{11}&=&{4\I m \over \Lambda}e^{\nu}r^2\varpi \\
e_{12}&=& -{\I m \over \Lambda}r^2 e^{-\lambda}\omega'\\
e_{13}&=& -e_{12} \\
e_{20}&=& -{4 \over \Lambda}e^{\nu}r^2 \varpi {{\cal L}_2^\pm} \\
e_{21}&=& -{\omega r \over \Lambda} {{\cal L}_2^\pm}
\end{eqnarray}

Background coefficients for the evolution equation
(\ref{eq:evol_H}) for $H$
\begin{eqnarray}
  k_{01} &=& \frac{r C_s^2}{\alpha+\beta} e^{\nu} \left(\beta-3\alpha-2re^{-\lambda}\right)\\
  k_{02} &=& \frac{r C_s^2}{\alpha+\beta} \Lambda e^{\nu}  \\
  k_{03} &=& \frac{C_s^2}{2r(\alpha+\beta)}(\beta-3\alpha+4M) \\
  k_{04} &=& \frac{r C_s^2}{2(\alpha+\beta)} e^{-\nu}(3\beta-\alpha+4M)\\
  k_{05} &=& -\frac{C_s^2}{\alpha+\beta}e^{-\lambda}(\beta-\alpha+2M)\\
  k_{06} &=& -\frac{C_s^2}{\alpha+\beta} r^3 e^{\nu-\lambda}\\
  k_{07} &=& \frac{\alpha}{r^2} +
  \frac{C_s^2}{r^2(\alpha+\beta)}\left[(\alpha-\beta)(2r-6M-\beta)-2\alpha\beta-4rMe^{-\lambda}\right]
\\
  k_{10} &=& \frac{2\I m}{r\Lambda}(\alpha+\beta)\left[(C_s^2-1)\omega+2\varpi C_s^2\right]
  +\I m\left( C_s^2\varpi-\Omega\right)
  \\
  k_{11} &=& \frac{2\I m}{\Lambda}e^{-\nu}C_s^2 \left[\frac{\Lambda r (\beta+M)}{\alpha+\beta}+\left(\alpha+\beta-re^{-\lambda}\right)\right]\omega
  \nonumber \\
&+&\frac{\I m}{2\Lambda}e^{-\nu}C_s^2   +\left[r\Lambda-4(\alpha+\beta)\right]\varpi
\\
  k_{12} &=& \frac{\I m}{r^2\Lambda} C_s^2 \left[r\Lambda(1-4e^{-\lambda})-3\beta-\alpha-4M\right]\omega
  \nonumber \\
  &-&\frac{\I m}{r^2\Lambda}C_s^2\left[2(4M-3\alpha+\beta) +\frac{r\Lambda(8M+3\beta-5\alpha)}{2(\alpha+\beta)}\right]\varpi  
   \\
  k_{13} &=& \frac{\I m}{r\Lambda}C_s^2 (\beta-\alpha)\omega \\
  k_{14} &=& -\frac{\I m}{\Lambda}C_s^2 re^{-\lambda}\omega \\
  k_{21} &=& \frac{\omega}{r^2\Lambda}\left[\Lambda \alpha {{\cal L}_1^\pm}-(\alpha-\beta)\Lambda^\pm C_s^2 {{\cal L}_2^\pm}\right]
  +\frac{\Lambda-2}{\Lambda}e^{-\lambda}C_s^2\omega' {{\cal L}_2^\pm}
  \nonumber \\
  &+&\frac{C_s^2}{r^3\Lambda}\left[r(\alpha-M)e^{-\lambda}-4(\alpha+\beta)^2\right]\varpi {{\cal L}_2^\pm}
  \nonumber \\
  &+&\left\{\frac{2(\alpha-\beta)}{r^2(\alpha+\beta)}(r-3M-\beta)+\frac{\beta}{r^2}-4rMe^{-\lambda}\right\}C_s^2\omega {{\cal L}_1^\pm}
  \nonumber \\
  &+& \left[ \frac{\alpha}{r^2}-\frac{3\beta}{r^2}C_s^2-\frac{2(\beta-3\alpha+4M)}{r(\alpha+\beta)}e^{-\lambda}C_s^2\right] \varpi {{\cal L}_1^\pm}
  \\
  k_{22} &=& \left[ 3{{\cal L}_1^\pm}+\frac{4}{r\Lambda}(\alpha+\beta){{\cal L}_2^\pm}\right]C_s^2e^{-\lambda}\varpi 
  \nonumber \\
  &+& \left[ \frac{\alpha+3\beta+2M}{\alpha+\beta}{{\cal L}_1^\pm}+\frac{\Lambda^\pm}{\Lambda}{{\cal L}_2^\pm}\right]C_s^2e^{-\lambda}\omega
\end{eqnarray}

Background coefficients for the evolution equation
(\ref{eq:evol_R}) for $R$
 \begin{eqnarray}
  f_{00} &=& \frac{1}{2r^4}\left(\beta-3\alpha+4M\right) \\
  f_{01} &=& -\frac{3}{2r^2}e^{-\nu}(\alpha+\beta) \\
  f_{02} &=& -\frac{\alpha+\beta}{r^3}\\
  f_{03} &=& -\frac{1}{2r^5} \left(\beta-3\alpha+4M\right)\\
  f_{04} &=& -\frac{1}{2r^4}e^{\nu}\left[(11\alpha+3\beta-8M)r-6\alpha e^{\lambda}(\alpha+\beta) \right] \\
  f_{10} &=&  -\frac{2\I m}{r^3}e^{-\nu}\varpi(\alpha+\beta)\\
  f_{11} &=&  -\frac{\I m}{r\Lambda}\left(16M-16\alpha-\Lambda
  r\right)\omega' \nonumber \\
  &-& \frac{2\I m}{r^2\Lambda}\left[\Lambda(\alpha e^{-\lambda}-r)
          +8(\beta-3M+4\alpha)\right]\varpi\\
  f_{12} &=& -\frac{\I m}{r\Lambda}
           \left[r\Lambda\Omega + 8\varpi(\beta+M)\right] \\
  f_{13}  &=& \frac{2\I m}{r^2\Lambda}e^{-\nu}(\beta-2\alpha+3M)\omega'
  \nonumber \\
         &+&\frac{2\I m}{r^4\Lambda}e^{-\nu}\left[ 5Mr+2\alpha e^{\lambda}(2\alpha-r)-r(\beta+4\alpha)\right]\omega
	 \\
  f_{14} &=&-\frac{2\I m}{r^2\Lambda} e^{-\nu}\left(\beta-M+2\alpha\right)\omega \\
  f_{15} &=&\frac{16\I m}{r\Lambda}(M-\alpha)\varpi\\
  f_{20} &=& \frac{e^{-\nu}}{r^4\Lambda}(\alpha+\beta)\left[32(M-\alpha){{\cal L}_2^\pm} +r\Lambda {{\cal L}_1^\pm}\right]\Omega
  \nonumber \\
  &+& \frac{4}{r^2\Lambda}e^{-\nu-\lambda}\left(\beta-\alpha+3M\right)\omega' {{\cal L}_2^\pm}
  \nonumber \\
  &-&\frac{2}{r^3\Lambda}e^{-\nu}(M+\beta)\omega {{\cal L}_2^\pm}
  \nonumber \\
  &+&\frac{2}{r^4\Lambda}e^{-\nu}\left[2(\alpha+\beta)(r\Lambda^\pm+8\alpha-8\beta)-r(M+\beta)\right]\omega{{\cal L}_2^\pm}
  \\
  f_{21} &=& \frac{1}{r^5}e^{-\nu}(\beta-\alpha+2M)\left[2e^{\lambda}\alpha\Omega-r^2\omega'\right] {{\cal L}_1^\pm}
  \nonumber \\
 &+& \frac{32}{r^5\Lambda}e^{-\nu}(\alpha+\beta)(\alpha-M)\Omega {{\cal L}_2^\pm}
 \nonumber \\
 &-&\frac{2}{r^5\Lambda}e^{-\nu}(\beta+M)
\left(3 r^2\omega'+2e^{\lambda}\alpha\omega\right){{\cal L}_3^\pm}
 \nonumber \\
 &-&\frac{4}{r^5\Lambda}e^{-\nu}\left[\alpha(M+\beta)e^{\lambda}+8(\alpha+\beta)(M-\alpha)\right]\omega {{\cal L}_2^\pm}
 \nonumber \\
 &-& \frac{4\Lambda^\pm}{r^5\Lambda}e^{-\nu}\left[(2\alpha^2+\beta\alpha-M\alpha)e^{\lambda}-2r(\alpha-M)\right]\omega {{\cal L}_2^\pm}
  \nonumber \\
  &+& \frac{2}{r^3\Lambda}e^{-\nu}\left[(4\alpha-\beta+3M)\Lambda^\pm +4\alpha-5\beta-9M\right]\omega'{{\cal L}_2^\pm}
  \nonumber \\
  &+& \frac{16}{r^4\Lambda}e^{-\nu}\left[2(\alpha-M)^2+M^2-\beta M+2\beta \alpha\right]\omega' {{\cal L}_2^\pm}
  \nonumber \\
  &+& \frac{16}{r^6\Lambda}e^{-\nu-\lambda}\varpi (\alpha+\beta)\left[3rMe^{-\lambda}+\beta(2M-r)+\alpha(3\alpha-4r+5M)\right]{{\cal L}_2^\pm}
  \nonumber \\
  &-& \frac{16}{r^6\Lambda}e^{-\nu-\lambda}\varpi \alpha (\alpha+\beta)(M-\alpha)C_s^{-2}{{\cal L}_2^\pm} 
  \nonumber \\
  &-&\frac{2}{r^5}e^{-\nu-\lambda}\varpi \left[\beta(\alpha-r)-\alpha(\alpha+r)+2M(\beta+2\alpha)\right]{{\cal L}_1^\pm}
  \\
  f_{22} &=& \frac{2}{r^2}\left(r-2\alpha e^{\lambda}\right)\varpi {{\cal L}_1^\pm} +\left[\frac{32}{\Lambda r}(\alpha-M){{\cal L}_2^\pm}-{{\cal L}_1^\pm} \right]\varpi'
  \nonumber \\
  &-&\frac{16}{r^2\Lambda}\left(4\alpha+\beta-3M\right)\omega{{\cal L}_2^\pm}
  \\ 
  f_{23} &=& \frac{16}{r\Lambda}\varpi (M-\alpha) {{\cal L}_2^\pm}\\
  f_{24} &=& \frac{3\Lambda^\pm}{r^5\Lambda} (\beta+M) (\beta-\alpha)\omega {{\cal L}_2^\pm}   \\
  f_{25} &=&-\frac{3\Lambda^\pm }{r^3\Lambda}e^{-\lambda}(\beta+M)\omega {{\cal L}_2^\pm}  \\
  f_{26} &=& \frac{1}{r^3\Lambda}e^{-\nu}(\beta+M) \left[(\Lambda^\pm-2){{\cal L}_2^\pm} -2\Lambda {{\cal L}_1^\pm} -2{{\cal L}_3^\pm}\right]\omega
  \nonumber \\
  &-&\frac{2}{r^3}e^{-\nu}(\alpha+\beta)\varpi {{\cal L}_1^\pm}
  -\frac{16}{r^4\Lambda}e^{-\nu}(\alpha+\beta)(M-\alpha)\varpi{{\cal L}_2^\pm}  
  \nonumber \\
  &-&\frac{4}{r^3\Lambda}e^{-\nu-\lambda}\left(\beta-2\alpha+3M\right)\omega'{{\cal L}_2^\pm}         
 \end{eqnarray}

Background coefficients for the evolution equation
(\ref{eq:evol_V}) for $V$
\begin{eqnarray}
q_{00}&=& -\frac{\alpha+\beta}{r^3}\\
q_{01}&=& \frac{1}{r^2}(\alpha+\beta)e^{-\nu} \\
q_{02}&=& \frac{1}{r^4}(\beta-3\alpha+4M)\\
q_{10}&=&\frac{\I m}{r^3 \Lambda}\left[2\beta(r-M)-2r(2M-3\alpha)e^{-\lambda}-\alpha(2M+5\beta)-\alpha^2\right]e^{-\nu}\varpi \nonumber \\
&+& \frac{\I m}{r^3\Lambda}\left[\beta(\beta-3M)-2(re^{-\lambda}-M)(\beta-\alpha)-4rMe^{-\lambda}\right]e^{-\nu}\varpi \nonumber \\
&+&\frac{\I m}{r \Lambda}e^{-\nu-\lambda}\varpi'
\\
q_{11}&=&\frac{\I m}{r\Lambda} e^{-\nu-\lambda}\left[2(M-\alpha)+(\alpha-\beta-2M)C_S^2\right]\varpi
\\
q_{12}&=& \frac{\I m}{\Lambda}\left(2\varpi-\Lambda\Omega+\Lambda C_s^2\varpi\right) \\
q_{13}&=&\frac{\I m}{\Lambda}e^{-\lambda}r^2\varpi' \nonumber \\
      &-&\frac{\I m}{\Lambda}\left[C_s^2(3\alpha-\beta+2re^{-\lambda})+2(\alpha-r
      e^{-\lambda})\right]\varpi \\
q_{14}&=& \frac{\I m}{2\Lambda r^2}e^{-\nu}\varpi\left[(\beta-3\alpha+4M)C_s^2+2(\alpha+3\beta+2M)\right]
\nonumber \\
 &+& \frac{\I m}{2\Lambda r^2}e^{-\nu}(3\beta-\alpha+4M)\omega \\
q_{15}&=&\frac{\I m}{2\Lambda}e^{-2\nu} \left[(3\beta-\alpha+4M)C_s^2+2(\beta-\alpha+2M) \right]\varpi
\nonumber \\
&-& \frac{\I m}{2\Lambda}e^{-2\nu}(\alpha+\beta)\omega
 \\
q_{16}&=& -\frac{\I m}{\Lambda} e^{-\lambda} C_s^2 r^2 \varpi\\
q_{20}&=&-\frac{\Omega}{\Lambda}{{\cal L}_2^\pm}+
\frac{\Omega-2\omega}{\Lambda}{{\cal L}_2^\pm}\\
q_{21}&=& \frac{\varpi}{r^3\Lambda}e^{-\nu}(\alpha+\beta)
\left[{{\cal L}_3^\pm}-(1+\Lambda^\pm){{\cal L}_2^\pm}\right]
\nonumber \\
&-&\frac{\omega}{r^3\Lambda}e^{-\nu}\left[(\alpha+\beta)\Lambda^\pm {{\cal L}_2^\pm}+(\beta-\alpha+2M)({{\cal L}_2^\pm}+{{\cal L}_3^\pm})\right]
\end{eqnarray}

Background coefficients for the evolution equation
(\ref{eq:evol_U}) for $U$
\begin{eqnarray}
s_{00}&=& \frac{2}{r^5}\left[\beta(\alpha+\beta)+2re^{-\lambda}(M-\alpha)\right]   \\
s_{01}&=& -\frac{2}{r^3}e^{-\lambda}(\alpha+\beta) \\
s_{10}&=& -\I m \Omega -\frac{2\I m}{r\Lambda}\left(4\beta-r +4M\right)\varpi \\
s_{11}&=& -\frac{2\I m}{r^4\Lambda}e^{-\nu}
(\alpha+\beta)\left(4\beta+r\Lambda+r-2re^{-\lambda}\right)\varpi
 \nonumber \\
 &-&\frac{4\I m}{r^2\Lambda}e^{-\nu-\lambda}r^2(\beta+M)\omega'   \\
s_{12}&=& \frac{2\I m}{r \Lambda} e^{-\nu-\lambda}(M+\beta)\omega' \\
s_{13}&=& -s_{12}\\
s_{20}&=& \frac{1}{r\Lambda}\left[(r+8M+8\beta-rC_s^\pm\Lambda^\pm)\varpi +r\omega  \right]{{\cal L}_2^\pm}
\nonumber \\
  &+&\frac{\omega-\varpi}{\Lambda}{{\cal L}_3^\pm}\\
s_{21}&=& \frac{1}{\Lambda}\left[r^2e^{-\lambda}\varpi' +2(\alpha
e^{-\lambda}-r)\varpi +C_s^2\left(3\alpha-\beta +2 r
e^{-\lambda}\right)\varpi\right] {{\cal L}_2^\pm}
\\
 s_{22}&=& \frac{\varpi e^{-\nu}}{r^3\Lambda}\left[(3\beta-2r+6M)\alpha-\beta^2 +2\beta(r-3M)+4rMe^{-\lambda} \right]C_s^2{{\cal L}_2^\pm}
 \nonumber \\
&+& \frac{\varpi e^{-\nu}}{r^3\Lambda}\left[(5\beta-6r+14M)\alpha+\alpha^2-2(r-M)\beta+4rMe^{-\lambda}\right]{{\cal L}_2^\pm}
 \nonumber \\
 &-&\frac{e^{-\nu-\lambda}}{r\Lambda}(\alpha+\beta){{\cal L}_2^\pm}
\\
s_{23}&=& \frac{e^{-\nu-\lambda}}{r\Lambda}\varpi
 \left[ (\beta-\alpha+2M)C_s^2+2(M-\alpha)\right]{{\cal L}_2^\pm} \\
 s_{24}&=& \frac{C_s^2}{\Lambda}r^2 e^{-\lambda}\varpi {{\cal L}_2^\pm}\\
 s_{25}&=&\frac{e^{-\nu}}{2r^2\Lambda}(\alpha+\beta)\Omega{{\cal L}_2^\pm}
 \nonumber \\
 &-& \frac{e^{-\nu}}{2r^2\Lambda}\left[(\beta-3\alpha+4M) C_s^2+(7\beta+3\alpha+4M) \right]\varpi{{\cal L}_2^\pm}
  \\
 s_{26}&=&\frac{e^{-2\nu}}{2\Lambda}\left[(\alpha+\beta)\Omega-(C_s^2+1)(3\beta-\alpha+4M)\varpi \right]{{\cal L}_2^\pm}
\end{eqnarray}

\end{document}